\documentclass[fleqn,usenatbib]{mnras}

\DeclareRobustCommand{\VAN}[3]{#2}
\let\VANthebibliography\thebibliography
\def\thebibliography{\DeclareRobustCommand{\VAN}[3]{##3}\VANthebibliography}

\usepackage{newtxtext,newtxmath}
\usepackage[T1]{fontenc}
\usepackage{graphicx}	
\usepackage{amsmath}	
\usepackage{bbold}
\usepackage{comment}
\usepackage{bm}
\usepackage{mathtools}
\usepackage{multirow}
\usepackage{natbib}

\newcommand{\mvir}[0]{M_\mathrm{vir}}
\newcommand{\cvir}[0]{c_\mathrm{vir}}

\newcommand{\rmax}[0]{r_\mathrm{max}}
\newcommand{\vmax}[0]{v_\mathrm{max}}

\newcommand{\rt}{r_\mathrm{t}}
\usepackage{booktabs}
\usepackage{subcaption}
\usepackage{xcolor}
\usepackage{soul}
\usepackage[normalem]{ulem}
\usepackage[export]{adjustbox}

\renewcommand{\vec}[1]{{\bm{#1}}}

\title[Dense perturbers in two lens systems]{SHARP -- IX. The dense, low-mass perturbers in B1938+666 and J0946+1006: implications for cold and self-interacting dark matter
}

\author[Tajalli et al.]
{M. Tajalli$^{1}$\thanks{E-mail: mtajalli@mpa-garching.mpg.de}, S. Vegetti$^{1}$, C.~M. O'Riordan$^{1}$, S. D.~M. White$^{1}$, C.~D. Fassnacht$^{2}$, D.~M. Powell$^{1}$,
\newauthor
J.~P. McKean$^{3,4,5}$ and G. Despali$^{6,7,8}$
\\
$^{1}$Max Planck Institute for Astrophysics, Karl-Schwarzschild-Stra\ss{}e 1, 85748 Garching bei M\"unchen, Germany\\
$^{2}$Department of Physics and Astronomy, University of California Davis, 1 Shields Avenue, 95616, Davis, CA, USA\\
$^{3}$Kapteyn Astronomical Institute, University of Groningen, PO Box 800, NL-9700 AV, Groningen, The Netherlands\\
$^{4}$South African Radio Astronomy Observatory, Krugersdorp 1740, P.O. Box 443, South Africa\\
$^{5}$Department of Physics, University of Pretoria, Lynnwood Road, 0083, South Africa\\
$^{6}$Dipartimento di Fisica e Astronomia "Augusto Righi", Alma Mater Studiorum Università di Bologna, via Gobetti 93/2, I-40129 Bologna, Italy\\
$^{7}$INAF-Osservatorio di Astrofisica e Scienza dello Spazio di Bologna, Via Piero Gobetti 93/3, I-40129 Bologna, Italy\\
$^{8}$INFN-Sezione di Bologna, Viale Berti Pichat 6/2, I-40127 Bologna, Italy}

\date{Accepted XXX. Received YYY; in original form ZZZ}

\pubyear{2025}

\begin{document}
\label{firstpage}
\pagerange{\pageref{firstpage}--\pageref{lastpage}}
\maketitle

\begin{abstract}
We present an extended analysis of the gravitational lens systems SDSS J0946+1006 and JVAS B1938+666. We focus on the properties of two low-mass dark matter haloes previously detected in these systems and compare them with predictions from different dark matter models. In agreement with previous studies, we find that the object ${\cal H}$ detected in J0946+1006 is a dark-matter-dominated subhalo. Object ${\cal A}$, in B1938+666, is a foreground halo at $z = 0.13\pm0.07$, contradicting previous analyses which suggested this object to be located either within or at higher redshift than the lens. Given the new redshift for this object, we update the 3$\sigma$ upper limit on its luminosity to $L_V < 6.3 \times 10^5 {(z/0.13)}^2 L_{V,\odot}$. By selecting central galaxies from the TNG50 hydrodynamical simulation, we find that analogues with projected mass density profiles around the robust radius of $\sim$ 91 pc and luminosities consistent with detection $\mathcal{A}$ can be found, although they lie near the edge of the halo distribution in the relevant mass and redshift ranges. We conclude, therefore, that this object is an atypical but possible event in $\Lambda$CDM.
The projected mass density profile of both detections over the well-constrained range of radii may be consistent with expectations from SIDM gravothermal fluid model if the effective self-interaction cross-section $\sigma_{c,0}/m_{\rm{dm}}$ is of order $300 \ \rm{cm}^2 g^{-1}$ or larger.\\

\end{abstract}



\section{Introduction}

The dark energy plus Cold Dark Matter ($\Lambda$-CDM) paradigm for the growth of structure in the Universe has long served as the standard model of cosmology, owing to its success in explaining large-scale structure formation within a relatively simple framework with only a few independent parameters \citep[e.g.,][]{Planck_Collaboration, Alam_2021}. However, the CDM model has faced persistent challenges on smaller scales, where discrepancies have emerged between low-redshift observations in the local Universe and theoretical predictions. Although several of these issues have been resolved or alleviated through advances in high-resolution hydrodynamical simulations that incorporate baryonic physics, some of the tensions remain unresolved. 
For instance, recent studies suggest that cluster tides are largely responsible for both the lack of low surface brightness (LSB) dwarf galaxies in the central regions of the Fornax Cluster and the disturbed morphologies observed in its LSB dwarf spheroidal galaxies \citep{Asencio_2022}. Under the assumption that these systems are dark-matter-dominated, this indicates a degree of tidal susceptibility that is incompatible with the $\Lambda$-CDM framework. This tension could be alleviated if these dwarfs had significantly lower dark matter fractions; however, as shown in \citet{Asencio_2022}, it is highly unlikely that baryonic feedback could have substantially altered their inner dark matter distributions. Moreover, while the diversity of galactic rotation curves \citep[RCs,][]{Oman_2015} continues to pose a challenge to CDM based on observations of the Local Group, a tight radial acceleration relation (RAR) identified in various samples of galaxies makes it particularly difficult to explain both the diversity of RC shapes and the RAR simultaneously within the CDM paradigm \citep{Famaey_2012, Lelli_2017, Banik_2022, Stiskalek_2023}. Nevertheless, the origin of the RAR and the mass-discrepancy acceleration relation (MDAR) remain a subject of debate in the literature, with some studies arguing that these can arise naturally from galaxy formation processes in a $\Lambda$-CDM cosmology \citep{Keller_2017, Ludlow_2017, Navarro_2017}.

These issues have primarily been identified in the context of near-field cosmology, but it is imperative to extend tests of the standard cosmological model beyond the local Universe. Strong gravitational lensing offers a promising tool to achieve this goal \citep[e.g.][and references therein]{Natarajan_2024, Vegetti_2024}. Specifically, strong gravitational lensing observations can be used to detect low-mass dark matter haloes, that would not be otherwise observable due to their low baryonic content, via their gravitational effect on the multiple magnified lensed images \citep[e.g.,][]{Koopmans_2005,Vegetti_2009,Mao_1998,Dalal_2002,Gilman_2019a}. As the number and properties of these objects strongly depend on the physics of dark matter \citep[e.g.,][]{Springel_2008, Lovell_2014, Colin_2002, Vogelsberger_2012}, this effect provides an  avenue to test different dark matter models \citep[e.g.,][]{Vegetti_2014,Vegetti_2018,Gilman_2019b,Hsueh_2020,Enzi_2021}. 

To date, two low-mass dark matter haloes have been gravitationally imaged \citep[i.e detected via pixellated corrections to the lensing potential,][]{Koopmans_2005,Vegetti_2009} in the optical imaging data of the strong gravitational lens systems SDSS J0946+1006 \citep[][detection ${\cal H}$]{Vegetti_2010} and JVAS B1938+666 \citep[][detection ${\cal A}$]{Vegetti_2012}. 

Both detections have since been independently confirmed in subsequent analyses. Their concentrations have been shown to be unexpectedly high, deviating from the mass-concentration relations derived from CDM $N$-body simulations \citep{Minor_2021, Seng_l_2022, Ballard_2024, Minor_2025, Despali_2024, Enzi_2024}.
From a comparison with hydrodynamical simulations \citet{Minor_2021} and \citet{Despali_2024} have concluded that there is no CDM analogue of detection ${\cal H}$, while detection ${\cal A}$ is less unusual when assumed to be a subhalo of the main lens galaxy. An additional subhalo detection has been reported by \citet{Hezaveh_2016} using ALMA observations of the gravitational lens system SPD81 but this has not so far been independently confirmed.

This paper presents a new analysis of the properties of detections ${\cal H}$ and ${\cal A}$  and how they compare with predictions from cold and self-interacting dark matter (SIDM). Our work improves upon previous studies in several aspects. Apart from \cite{Despali_2024}, ours is the only analysis focusing on both objects, and in comparison to their work, we allow a more complicated mass model for the lens galaxy, going beyond a single power-law model and allowing extra angular complexity in the form of multipoles of order one, three and four. We also explicitly test the effect of different priors for the background source galaxy. As in \citet{Seng_l_2022} and \citet{Enzi_2024}, we treat the redshift of the low-mass haloes as a free parameter. We also test different assumptions for their mass density profiles. While \citet{Seng_l_2022} analyzed the \textit{Hubble} Space Telescope (HST) NICMOS-1 data for  B1938+666, we focus on the higher-resolution, higher-SNR Keck-II adaptive optics (AO) data obtained by the Strong Lensing at High Angular Resolution Programme (SHARP).

The paper is organized as follows. Sections \ref{sec:data} and \ref{sec:method} introduce the data and how they were modelled, respectively. We present the results of our analysis in Section \ref{sec:modelling_results}. We compare the inferred properties of detections ${\cal A}$ and ${\cal H}$ with predictions from CDM hydrodynamical simulations, specifically with TNG50 \citep{Pillepich_2019, Nelson_2019} in Section \ref{sec:compare_with_cdm}, and with SIDM gravothermal fluid models \citep{Outmezguine_2023} in Section  \ref{sec:compare_with_sidm}. We summarise and discuss our findings in Section \ref{sec:summary}.

\section{Data}
\label{sec:data}

\begin{figure*}
    \centering
    \includegraphics[width=\textwidth]{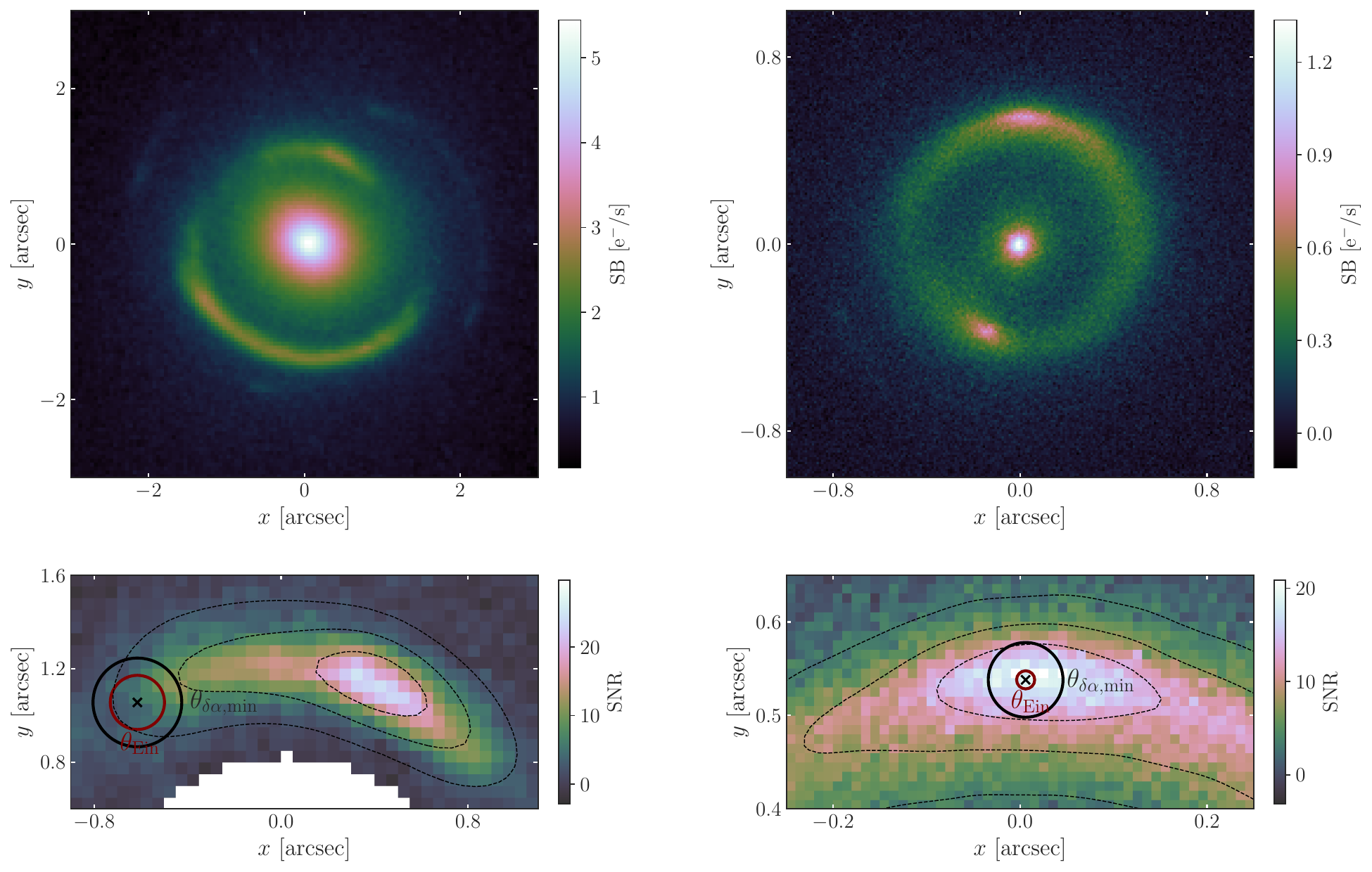}
    \caption{Top panels: F814W HST ACS data for SDSS J0946+1006  (left) and \textit{K$'$} band Keck-II AO data for the gravitational lens system JVAS B1938+666 (right). Bottom panel: the colour maps and contours represent the data SNR. The crosses mark the apparent positions of detections ${\cal H}$ (left) and ${\cal A}$ (right). The red circles indicate the Einstein radii, inferred from the PL model, and the black circles are the robust radii, where the relative error on the deflection angle is minimised for the best-fitting models (see Section \ref{sec:robust_radius}).}
    \label{fig:data_and_snr}
\end{figure*}

The data used to model the SDSS J0946+1006 lens system were obtained in the F814W \textit{I}-band with the Advanced Camera for Surveys (ACS) instrument aboard the HST \citep{Gavazzi_2008}. The system consists of three background sources at redshifts $z_{\rm{s1}} = 0.609$, $z_{\rm{s2}} = 2.035$ \citep{Smith_2021}, and $z_{\rm{s3}} = 5.975$ \citep{Collett_2020}, lensed by a foreground galaxy at  $z_{\rm{lens}} = 0.222$. In this paper, we only model the lensed images of the lowest-redshift source, i.e., the brighter inner arcs as shown in Fig. \ref{fig:data_and_snr}.

JVAS B1938+666 was first reported as a gravitational lens in multi-frequency radio observations \citep{King_1997}, and its highly-magnified infrared Einstein ring was later detected in the F160W \textit{H}-band using NICMOS on board the HST \citep{King_1998}. The system has a main lens galaxy at redshift $z_{\rm{lens}} = 0.881$ \citep{Tonry_2000} and a background source at $z_{\rm{s}} = 2.059$ \citep{Riechers_2011}.
High-resolution imaging of the lens system was later obtained with the Keck-II telescope as part of SHARP, using the Near Infrared Camera 2 (NIRC2) adaptive optics in the \textit{H} and \textit{K$'$} bands \citep{Lagattuta_2012, Vegetti_2012}. 
While our main analysis relies on the higher-quality \textit{K$'$} band Keck-II AO data, we also model the \textit{drizzled} HST NICMOS-1 observations for comparison with \citet[][see Appendix \ref{app:NIC1}]{Seng_l_2022}.

\section{Method}
\label{sec:method}

The data analysed in this work are modelled with the \textsc{pronto} lens modelling code, a semi-linear inversion method embedded in a hierarchical Bayesian framework with an adaptive, resolved background source \citep{Vegetti_2009,Rybak_2015,Rizzo_2018,Ritondale_2019,Powell_2021, Ndiritu_2024}. 

\subsection{Main lens}

\subsubsection{Mass density profile}

We model the mass distribution of the lens galaxy with a singular elliptical power-law (PL) with dimensionless surface mass density, i.e., convergence, given by
\begin{equation}
\kappa(R) = \kappa_0\left(1-\frac{\alpha}{4}\right)\frac{q^{\alpha-3/2}}{(R/r_{\rm p})^{(\alpha-1)}}\,,
\label{eq:ple}
\end{equation}
where $R^2 = (qx)^2+y^2$ is the elliptical radius, $\kappa_0$ is the surface mass density normalization, $\alpha$ is the 3D logarithmic density slope ($\alpha =2$ for an isothermal profile), $q$ is the projected axis ratio, $r_p\equiv$ 1 arcsec is the pivot radius. Additionally, we include an external shear component of strength $\Gamma$ and position angle $\theta_\Gamma$. 
We account for deviations from perfect ellipticity in the mass distribution of the lens galaxy by adding multipoles of order $m= 1, 3, 4$. These are Fourier-like corrections to the main density profile. The convergence of the multipole terms, expressed in polar coordinates, can be written as 
\begin{equation}
\kappa_m(R^\prime, \theta) = \kappa_0 \left(\frac{R^\prime}{r_{\rm p}}\right)^{-\left(\alpha - 1\right)}\left[ a_m \sin(m\theta) + b_m \cos(m\theta) \right]\,,
\label{eq:multipoles_spherical}
\end{equation}
where $a_m$ and $b_m$ are dimensionless coefficients, and quantify the strength and position angle of multipole perturbations. $R^\prime$ is the circular radius and $\alpha$ is the same density slope for the elliptical power-law profile as defined in Eq. (\ref{eq:ple}).

We assume uniform priors over a range of $\pm$1 per cent for multipole coefficients of order $m=3$ and $m=4$, and $\pm 10$ per cent for $m=1$. These choices are motivated by previous work on the strength of multipoles in the total mass density distribution of lens galaxies \citep{Kochanek_2004, Powell_2022, Stacey_2024} and early-type galaxies in general \citep{Amvrosiadis_2024}.

\subsubsection{Light profile}

To assess the robustness of the inferred properties of detections ${\cal H}$ and ${\cal A}$, we analyze each dataset under two different assumptions for the lens light profile. 

First, we model the lens light progressively by adding one elliptical Sérsic component at a time and fitting them simultaneously with the lens and perturber mass distributions within \textsc{pronto}. We do so until the Bayesian evidence begins to decline, indicating that further components are not supported by the data. 
To complement the modelling within \textsc{pronto}, we also model the lens light independently using the \textsc{Galfit} fitting algorithm \citep{Peng_2010} and following a similar strategy: additional complexity is included until further additions no longer result in a meaningful improvement in the reduced chi-squared or in the residual structure. The best-fit light profiles are then subtracted from the data prior to lens modelling.
Both light modelling approaches agree and result in three Sérsic components for SDSS J0946+1006 and two for JVAS B1938+666.

\subsection{Low-mass haloes}

We model the mass density profiles of each detected low-mass object using the following spherical models.

The first is the NFW profile \citep{Navarro_1996}, for which the 3D density is given by
\begin{equation}
\label{eq:nfw}
\rho(r) = \frac{\rho_s}{(r/r_s) \left( 1 + r/r_s \right)^2}\,,
\end{equation}
where $\rho_{\rm{s}}$ is the density normalization and $r_{\rm s}$ is the scale radius at which the 3D logarithmic slope of the density profile is $-2$, the slope of a singular isothermal sphere. 

We use different parametrisations of this profile for field haloes and for subhaloes. Field haloes are characterized by their virial mass, $M_{\rm{vir}}$, and their concentration, $c_{\rm{vir}} = R_{\rm{vir}}/r_s$, where $R_{\rm{vir}}$ is the virial radius, within which the mean density is $\Delta_{\rm{vir}}$ times the mean density of the Universe at the redshift of the halo \citep{Bullock_2001}{}. Subhaloes lack a well-defined virial mass or radius, so we parametrize their NFW profile in terms of the more relevant quantities $v_{\rm{max}}$, their maximum circular velocity, and $r_{\rm{max}}$, the radius at which this maximum is achieved. For an NFW profile, these are related to the scale radius and the characteristic density through $r_{\rm max} = 2.163 \, r_s$ and $v_{\rm max} = 1.64 \, r_s \sqrt{G\rho_s}$ \citep[e.g][]{Klypin_2015}. We can then use the parameter $c_\mathrm{V}$ as an alternative definition for the concentration
\begin{equation}
c_\mathrm{V} = 2 \left( \frac{\vmax}{H(z)\, \rmax} \right)^2\,,
\end{equation}
where $H(z)$ is the Hubble parameter at redshift $z$ \citep{Springel_2008}. We model NFW field haloes either with five free parameters, namely projected position, $\mvir$, $\cvir$, and redshift, or with four, taking $\cvir$ to be the median of the mass–concentration–redshift relation from \citet{Duffy_2008}. NFW subhaloes are modelled with four free parameters, the projected position,  $\vmax$ and $\rmax$; their redshift is taken to be that of the main lens.

Our second profile is the Pseudo-Jaffe (PJ) profile \citep{Munoz_2001}, defined as 
\begin{equation}  
\rho(r) = \frac{\rho_0 r_t^4}{r^2 (r^2 + \rt^2)}\,,
\end{equation}
where $\rho_0$ is the density normalisation and $\rt$ is the truncation radius, beyond which the density drops steeply as $r^{-4}$. This profile has a total of five degrees of freedom, including the redshift and position on the sky.

In addition to the NFW and PJ profiles, we consider a spherical power-law density distribution, defined as 
\begin{equation}
    \rho(r) = \rho_{\rm p} \left( \frac{r}{r_{\rm p}} \right)^{-\gamma}\,,
\end{equation}
where $\rho_{\rm p}$ is the density normalization and $\gamma$ is the 3D logarithmic slope. For reasons explained later in the paper (see Section \ref{sec:modelling_results}) the pivot radius is set to $r_{\rm p} = D_\mathrm{a}(z)\theta_p$ where $D_\mathrm{a}(z)$ is the angular diameter distance to the detected object and $\theta_p=0.19$ and 0.04 arcsec for detections $\cal{H}$ and $\cal{A}$, respectively. This leaves five free parameters in the model.

In addition to the profiles described above, we tested a two-component model consisting of an NFW profile and a central point mass representing a candidate black hole. As this model is disfavoured by both datasets, we do not consider it further in this paper. For all models, when the redshift is a free parameter of the model, we fit for the apparent positions of the low-mass haloes on the image plane rather than their true positions on their respective redshift planes. 

\section{Lens modelling results}
\label{sec:modelling_results}

The inferred parameters of objects $\cal{H}$ and $\cal{A}$ are given in Tables \ref{tab:J0046_pert} and \ref{tab:B1938_pert} for all considered mass models.

The statistical significance of each model $\cal{M}$ is the log-evidence difference relative to the reference model — that with the highest Bayesian evidence in each case, $\Delta\log{\cal{E}}_{\cal{M}} = \log{\cal{E}}_{\cal{M}} - \log{\cal{E}}_{\mathrm{ref}}$. The Bayesian evidence of model $\cal{M}$, defined as the probability of observing data $\vec{d}$ marginalised over all possible values of the model parameters, $\vec{\theta}$, is calculated as:
\begin{equation}
\label{eq:evidence}
\mathcal{E}_{\mathcal{M}} = \int P(\vec{d} \mid \vec{\theta}, \mathcal{M}) \, P(\vec{\theta} \mid \mathcal{M}) \, d\vec{\theta}.
\end{equation}
In all cases, the priors on the macro-model parameters remain the same. We evaluate the marginal likelihood in Eq. (\ref{eq:evidence}) using the MultiNest importance nested-sampling algorithm \citep{feroz13}. The posterior distributions for all models are given in Appendix~\ref{app:posteriors}. In the rest of the paper, we refer to a smooth power-law model (plus external shear) for the main lens as PL$_{\rm lens}$, and as PL$_{\rm lens}$+$\rm MP_{i,j,\ldots}$ when multipoles of orders i, j, etc., are included. Similarly, models including objects $\cal H$ or $\cal A$ will be referred to as PL$_{\rm lens}$+$\cal H$ (PL$_{\rm lens}$+$\cal A$) and PL$_{\rm lens}$+$\rm MP_{i,j,\ldots}$+$\cal H$ (PL$_{\rm lens}$+$\rm MP_{i,j,\ldots}$+$\cal A$). 

When modelling lenses in \textsc{pronto}, the source reconstruction can be changed in two important ways. First, the source regularisation (prior) can penalise against either strong gradients, or strong curvature in the surface brightness distribution\footnote{The regularisation matrices are built using discrete gradient and curvature operators defined on a triangular mesh \citep{Powell_2022}, an improvement on the finite-difference-based operators used in e.g. \cite{Vegetti_2009}.}. Second, we can choose the fraction of image plane pixels which are included in the linear solution for the source surface brightness. The values in pixels which are not included are interpolated from their neighbours. This is useful to avoid overfitting in systems with high angular resolution, where the number of pixels entering the source reconstruction is large.

For J0946+1006, we include all pixels within a mask on the image plane in the linear reconstruction, and adopt an adaptive regularisation scheme through weighting the regularisation constant by the inverse of the data signal-to-noise ratio. For B1938+666, we include only one in every four image plane pixels in the reconstruction when running the models presented in this work. We have verified that casting different number of pixels back to the source plane does not affect any of our main conclusions.

In this section and throughout the rest of the paper, we focus on results from gradient regularisation for J0946+1006 and from curvature regularisation for B1938+666. This is because we found in each case that  this choice consistently led to higher Bayesian evidence and improved image residuals in the final best-fitting model - independent of whether the perturbers were included and of which perturber profile was used. In Appendix \ref{sec:regularisation}, we discuss how different choices of source regularisation affect our results both in terms of the significance and properties of the two detections.

We find that whether or not the lens light is modelled simultaneously with the mass distribution, the properties of detection $\cal{A}$ remain consistent across all mass models, with only a marginal decrease of 0.8 in the log Bayes factor of the best-fit model. However, due to the more complex light profile of the foreground galaxy in the J0946+1006 lens system, for detection ${\cal H}$ we present results based only on models that simultaneously fit the lens light and mass distributions. This approach mitigates potential biases in the inferred perturber properties that may arise from degeneracies between the lens light, the reconstructed source and the low-mass halo.

\begin{table*}    
\caption{Median inferred parameters of detection $\cal {H}$ for different assumptions on its mass density profile. $\Delta\log{\cal E}$ is the log Bayes factor of each PL$_{\rm lens}$+$\rm MP_{1,3,4}$+$\cal H$ model relative to the best-fitting one, which is favoured over the PL$_{\rm lens}$+$\rm MP_{1,3,4}$ model by a log-evidence difference of 49.3. The projected enclosed mass and slope are measured at the system's robust radius of $\theta_{{\cal H}} = 0.19$ arcsec (see Section~\ref{sec:robust_radius}). The pivot
radius of the PL model is set to the robust angular scale, i.e., $r_{\rm p} =  D_\mathrm{a}(z)\theta_{\cal H}$. The uncertainties correspond to the 16th and 84th percentiles of the posterior distribution.}

\begin{tabular}{@{}cccccccc@{}}
\toprule
 $\mathrm{Profile}$ & $\Delta\log{\cal{E}}$ & $z$ & \multicolumn{2}{c}{Mass profile parameters} & & $M_{\rm{2D}}(\theta_{{\cal H}}) [10^9 M_\odot]$&$\gamma_{\rm{2D}}(\theta_{{\cal H}})$\\
\midrule
& & & $\vmax [\mathrm{km/s}]$ & $\rmax[\mathrm{kpc}]$ & $c_V [\times10^6]$ & \\
\midrule
\\
Subhalo NFW (free $\rmax$) & 0 & $\equiv 0.222$ & $123^{+6}_{-6}$
 & $0.75^{+0.19}_{-0.15}$ & $9.4^{+5.9}_{-3.7}$ & $3.44^{+0.33}_{-0.33}$ & $1.13^{+0.07}_{-0.07}$\\
\addlinespace[5pt] 
\midrule
\addlinespace[5pt]
& & & $M_{\rm t} [10^{9}\mathrm{M}_{\odot}]$ & $\log_{10}r_{\rm t}$ [kpc] & &\\
\midrule
\\
PJ (free $r_\mathrm{t}$) & $-$1.8 & $0.237^{+0.016}_{-0.014}$& $5.70^{+0.64}_{-0.64}$
 & $0.33^{+0.14}_{-0.11}$ & & $3.48^{+0.36}_{-0.34}$ & $1.19^{+0.06}_{-0.05}$\\
\addlinespace[5pt] 
\midrule
\addlinespace[5pt]
& & & $\mvir [10^{10}\mathrm{M}_{\odot}]$ & $\cvir$ & & \\
\midrule
\\
NFW (free $\cvir$) & $-$2.7 & $0.243^{+0.018}_{-0.016}$& $1.78^{+0.36}_{-0.32}$
 & $250^{+78}_{-53}$ & &  
 $3.39^{+0.38}_{-0.34}$ & $1.26^{+0.09}_{-0.09}$\\
\addlinespace[5pt] 
\midrule
\addlinespace[5pt]
&  & & $\rho_{\rm p} [10^{10}\mathrm{M}_{\odot}\rm{kpc}^{-3}]$ & $\gamma$ & & \\
\midrule
\\
PL & $-$5.0 & $0.239^{+0.018}_{-0.014}$& $0.017^{+0.004}_{-0.004}$ & $2.24^{+0.09}_{-0.08}$ & &  $3.16^{+0.38}_{-0.29}$ & $1.24^{+0.09}_{-0.08}$\\
\addlinespace[5pt] 
\midrule
\addlinespace[5pt]
& & & $\mvir [10^{12}\mathrm{M}_{\odot}]$ & $\cvir$ & \\
\midrule
\\
NFW (fix $\cvir$) & $-$30.1 & $0.243^{+0.020}_{-0.014}$& $5.92^{+0.69}_{-0.95}$
 & $\equiv6.36^{+0.11}_{-0.10}$ & &  
 $1.61^{+0.21}_{-0.16}$ & $0.215^{+0.004}_{-0.003}$\\
\addlinespace[5pt] 
\midrule
\addlinespace[5pt]
\end{tabular}
\label{tab:J0046_pert}
\end{table*}

\begin{table*}    
\caption{Median inferred parameters of detection $\cal {A}$ for different assumptions on its mass density profile. $\Delta\log{\cal E}$ is the log Bayes factor of each PL$_{\rm lens}$+$\rm MP_{3,4}$+$\cal A$ model relative to the best-fitting one, which is favoured over the PL$_{\rm lens}$+$\rm MP_{3,4}$ model by a log-evidence difference of 56.4. The projected enclosed mass and slope are measured at the system's robust radius of $\theta_{{\cal A}} = 0.04$ arcsec (see Section~\ref{sec:robust_radius}). The pivot
radius of the PL models is set to the robust angular scale, i.e., $r_{\rm p} = D_\mathrm{a}(z) \theta_{{\cal A}}$. The uncertainties correspond to the 16th and 84th percentiles of the posterior distribution.}
\begin{tabular}{@{}ccccccccc@{}}
\toprule
 $\mathrm{Profile}$ & $\Delta\log{\cal{E}}$ & $z$ & \multicolumn{2}{c}{Mass profile parameters} & & $M_{\rm{2D}}(\theta_{{\cal A}}) [10^7 M_\odot]$&$\gamma_{\rm{2D}}(\theta_{{\cal A}})$ & $M_{\rm{2D}}(\theta_{{\cal A}})/r_{\rm p} [10^5 M_\odot\rm{pc}^{-1}]$\\
\midrule
&  & & $\rho_{\rm p} [10^{10}\mathrm{M}_{\odot}\rm{kpc}^{-3}]$ & $\gamma$ & \\
\midrule
\\
PL & 0 & $0.123^{+0.073}_{-0.062}$ & $0.034^{+0.034}_{-0.016}$ & $1.84^{+0.13}_{-0.12}$ & & $3.07^{+1.68}_{-1.48}$ & $0.84^{+0.13}_{-0.12}$ & $3.41^{+0.28}_{-0.29}$\\
\addlinespace[5pt] 
\midrule
\addlinespace[5pt]
& & & $\mvir [10^8\mathrm{M}_{\odot}]$ & $\cvir$ & & \\
\midrule
\\
NFW (free $\cvir$) & $-$3.9 & $0.123^{+0.080}_{-0.056}$ & $5.08^{+3.86}_{-2.20}$
 & $185^{+116}_{-65}$ & & $3.02^{+1.74}_{-1.31}$ & $0.85^{+0.18}_{-0.14}$ & $3.36^{+0.30}_{-0.30}$\\
\addlinespace[5pt] 
\midrule
\addlinespace[5pt]

& & & $\mvir [10^{11}\mathrm{M}_{\odot}]$ & $\cvir$ &\\
\midrule
\\
NFW (fix $\cvir$) & $-$14.0 & $0.315^{+0.074}_{-0.049}$ & $3.74^{+2.28}_{-1.53}$
 & $\equiv7.60^{+0.20}_{-0.15}$ & & $6.07^{+1.01}_{-0.79}$ & $0.205^{+0.015}_{-0.012}$ & $3.21^{+0.25}_{-0.28}$\\
\addlinespace[5pt] 
\midrule
\addlinespace[5pt]

&  &  & $\rho_{\rm p} [10^{10}\mathrm{M}_{\odot}\rm{kpc}^{-3}]$ & $\gamma$ & & \\
\midrule
\\
PL & $-$17.7 & $\equiv 1.42$ & $0.085^{+0.029}_{-0.031}$ & $1.41^{+0.14}_{-0.11}$ & & $36.9^{+5.9}_{-5.9}$ & $0.41^{+0.14}_{-0.11}$ & $10.6^{+1.7}_{-1.7}$\\
\addlinespace[5pt] 
\midrule
\addlinespace[5pt]
&  & & $\rho_{\rm p} [10^{10}\mathrm{M}_{\odot}\rm{kpc}^{-3}]$ & $\gamma$ & & \\
\midrule
\\
PL & $-$23.8 & $\equiv 0.881$ & $0.014^{+0.011}_{-0.008}$ & $1.68^{+0.25}_{-0.18}$ & & $10.1^{+1.3}_{-1.3}$ & $0.68^{+0.25}_{-0.18}$ & $3.17^{+0.41}_{-0.41}$\\
\addlinespace[5pt] 
\midrule
\addlinespace[5pt]
\end{tabular}
\label{tab:B1938_pert}
\end{table*}

\subsection{SDSS J0946+1006: properties of detection \texorpdfstring{$\mathcal{H}$}{H}}

The most highly favoured model for detection $\cal{H}$ is one where its redshift is fixed at $z\equiv z_{\rm lens}$ and the inferred redshift for all other models lies within $\sim$1$\sigma$ from that of the main deflector. This result is consistent with the recent work of \cite{Enzi_2024} and confirms that this object, independently of its mass density profile, is best explained as a subhalo of the main lens galaxy. 
The increase in log-evidence for the best PL$_{\rm lens}$+$\rm MP_{1,3,4}$+$\cal H$ model relative to the PL$_{\rm lens}$+$\rm MP_{1,3,4}$ is 49.3 and this increase is still 19.2 for the worst fitting of the perturber models we tried. These increases in evidence become 75.6 and 31.5, when comparing the PL$_{\rm lens}$+$\cal H$ models against the baseline PL model. Hence, even though the statistical significance of this detection is sensitive to the presence of multipoles in the primary lens, the detection itself is robust. 

The subhalo model with the highest Bayesian evidence relative to all other models is an NFW profile with free $\rmax$, yielding a peak circular velocity of $\vmax = 123^{+6}_{-6}$ km s$^{-1}$, reached at a radius of $\rmax = 0.75^{+0.19}_{-0.15}$ kpc.
This model is then followed by the following field halo models: a PJ with free truncation radius, an NFW with free concentration, and a PL. The inferred virial mass and concentration of the NFW field halo are $\mvir = 1.78^{+0.36}_{-0.32} \times 10^{10} M_\odot$ and $\cvir=250^{+78}_{-53}$, respectively. The PL field halo model has a super-isothermal slope of $\gamma = 2.24^{+0.09}_{-0.08}$. Finally, the least-favoured model, with a substantial drop in evidence, $\Delta\log{\cal{E}} = -30.1$ relative to the NFW subhalo, is an NFW profile with concentration set by the median mass–concentration-redshift relation of \cite{Duffy_2008}. While all other models constrain the position of the perturber consistently, with maximum absolute deviations of $\Delta_{\rm{max}} x = 0.01$ and $\Delta_{\rm{max}} y = 0.01$ arcsec from the mean, the position of the perturber in this case is displaced by $\Delta x \sim 0.27$ and $\Delta y \sim 0.02$ arcsec, corresponding to nearly 5 pixels on the image plane. A similar shift in the inferred position of an NFW subhalo with fixed concentration was also reported by \cite{Despali_2024}. This behavior is likely a consequence of changes in the deflection angle field for a less concentrated and more massive halo, as shown in Fig. \ref{fig:projected_profs} (note that the deflection angle is related to the enclosed projected mass by $\alpha(r)\propto M_{\rm 2D}(r)/r$). The perturber must be placed farther away to reproduce the same lensing effect on the high-SNR regions in the data as other profiles. Our analysis confirms previous studies that concluded that detection $\cal{H}$ has an unexpectedly high concentration \citep{Minor_2021, Minor_2025, Despali_2024,Enzi_2024}. 

\subsection{JVAS B1938+666: properties of detection \texorpdfstring{$\mathcal{A}$}{A}}

The PL$_{\rm lens}$+$\rm MP_{3,4}$+$\cal A$ models we have tried have log-evidence values ranging from 56.4 down to 42.4 relative to the PL$_{\rm lens}$+$\rm MP_{3,4}$ model, while PL$_{\rm lens}$+$\cal A$ models yield log-evidence differences ranging from 58.2 down to 46.6 relative to the baseline PL$_{\rm lens}$. As discussed in Section \ref{sec:macro}, the $m=1$ multipole term is strongly disfavoured by the data for this lens system and is therefore excluded from the models presented below.

The model for $\cal{A}$ with the highest Bayesian evidence is a PL \textit{foreground} halo at a median redshift of $z = 0.12 ^ {+0.07}_{-0.06}$, with all other models inferring redshifts lower than that of the main lens and consistent with the best model within $1$ to $2\sigma$. We have tested the robustness of this result against different prior choices for the redshift (uniform in comoving distance and comoving volume), the presence of multipoles and the choice of lens light model. In all cases, the inferred redshift remains consistent within $1 \sigma$. A model where the redshift is forced to be that of the main lens, i.e. $z \equiv 0.881$, is disfavoured by the data with a $\Delta\log{\cal{E}} = {-}23.8$, relative to the best-fitting model. We, therefore, conclude that detection $\cal{A}$ is a foreground halo along the line of sight. These results contradict those by \citet{Seng_l_2022} who, from the lower-quality HST-NICMOS 1 data, have inferred a higher redshift of $z=1.42^{+0.10}_{-0.15}$. From the Keck-AO data analysed here, we find that a model with $z \equiv 1.42$ results in a large drop in evidence, $\Delta\log{\cal{E}} = {-}17.7$. We provide a more in-depth comparison with their analysis in Appendix \ref{app:NIC1}.

Our best fit PL foreground halo has sub-isothermal logarithmic slope $\gamma = 1.84^{+0.13}_{-0.12}$. The next best-fitting model is an NFW field halo with free concentration and redshift $z = 0.12^{+0.08}_{-0.06}$. We find a virial mass of $\mvir = 5.08^{+3.86}_{-2.20} \times 10^{8} M_\odot$ and a concentration $\cvir=185^{+116}_{-65}$.
Lastly, the worst-performing model for detection $\cal{A}$ is an NFW profile with concentration fixed to the median of the mass–concentration-redshift relation \citep{Duffy_2008}. This is strongly disfavoured by the data, as indicated by the drop in evidence, $\Delta\log{\cal{E}} = -14.0$. In this case, the inferred redshift and virial mass of the perturber are $z = 0.31^{+0.07}_{-0.05}$ and $\mvir = 3.74^{+2.28}_{-1.53} \times 10^{11} M_\odot$, respectively, indicating a shift in redshift towards the main lens and an increase in virial mass by nearly three orders of magnitude relative to the NFW model with a free concentration. This result is in agreement with Tajalli et al. (in prep) which shows that when a perturber moves away from the main lens towards the observer, its concentration must increase steeply to maintain a comparable lensing effect. However, imposing a mass–concentration-redshift relation suppresses the required increase in concentration, which is compensated by a higher virial mass and by placing the object at a redshift closer to the main lens. We note that the position of the detection is tightly constrained and consistent across all models, with maximum absolute deviations of $\Delta_{\mathrm{max}} x = 0.002$ and $\Delta_{\mathrm{max}} y = 0.002$ arcsec. The sub-isothermal slope found for $\mathcal{A}$ in all these models implies that the PJ profile (for which $\gamma\ge 2$ at all radii) is not a suitable model for its mass distribution. We therefore exclude this model from our analysis in this case.

It is important to emphasise that the particular parameters (e.g. $\mvir$ and $\cvir$) so far used to characterise  the mass profiles may not necessarily be robustly constrained because of possible strong degeneracies between them. In most cases, more careful parametrisation can remove or greatly reduce these degeneracies. In practice, the lensing effect of low-mass haloes can only be reliably measured over a limited radial range. In the following section, we focus on the projected mass density profile on scales where it is most accurately determined from the data.

\subsection{Projected quantities and robust radius}
\label{sec:robust_radius}

\begin{figure*}
\includegraphics[width=1.0\textwidth]{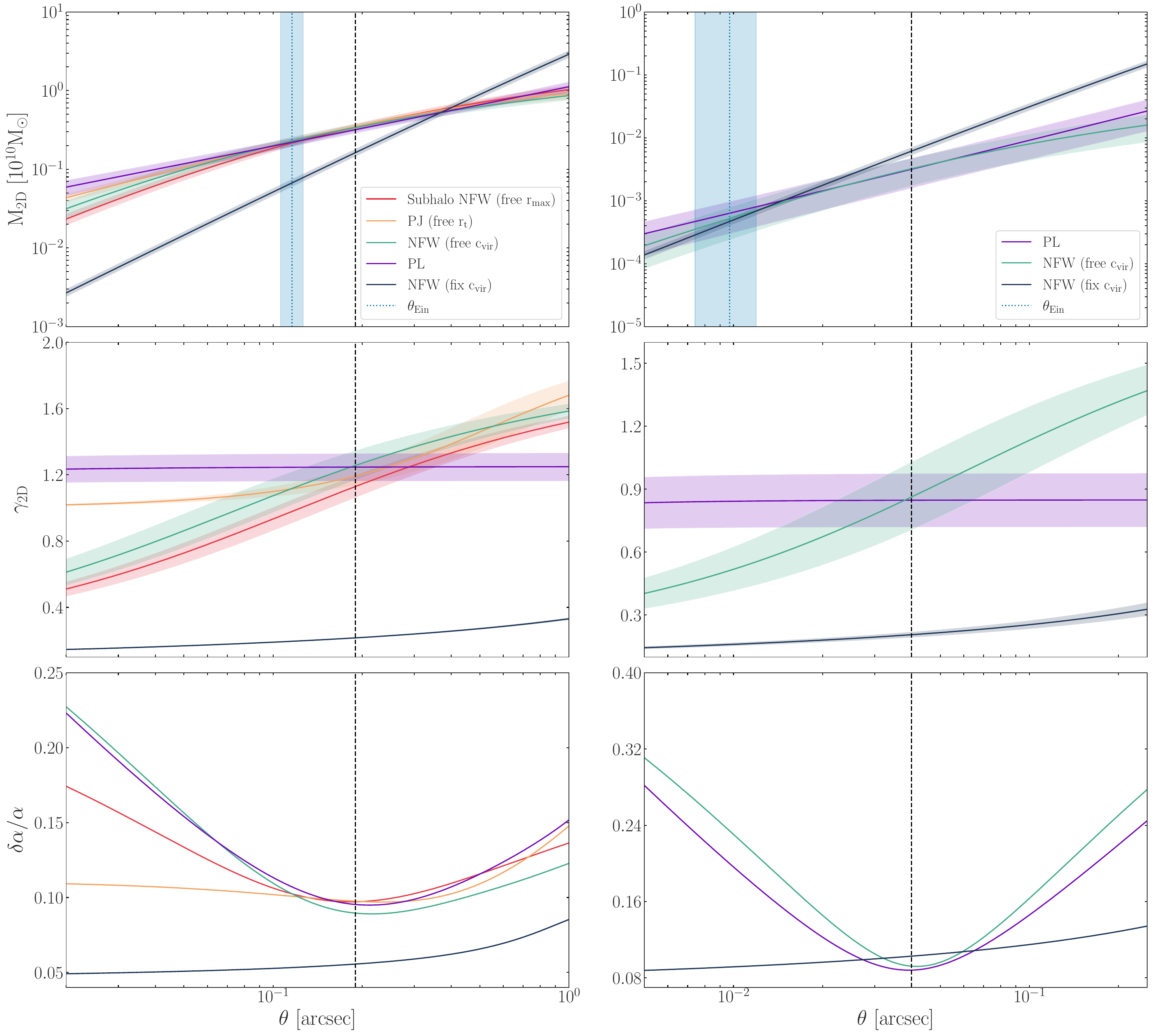}
\caption{Projected properties for detection $\mathcal{H}$ (left) and detection $\mathcal{A}$ (right). Top panels: the physical projected mass for the different parameterizations. The blue dashed line marks the Einstein radius as inferred from the PL model, with its 1$\sigma$ uncertainty shown as the blue shaded region. Middle panels: the logarithmic slope of the same profiles. Bottom panels: the relative error on the deflection angle. The deflection angle is calculated in the redshift plane of each model using the recursive form of the lens equation. In all panels, the black dashed line indicates the robust radius for each system.}
\label{fig:projected_profs}
\end{figure*}

Fig. \ref{fig:projected_profs} shows the projected mass density profiles and relative deflection angle errors for all mass models considered for detections ${\cal H}$ and ${\cal A}$, respectively.
 
We identify the radius at which the relative error on the deflection angle (and hence also on the projected mass) is minimised for the best-fitting model as the \emph{robust radius}, where the lensing effect of the low-mass halo is most reliably measured. We find this robust scale for detections ${\cal H}$ and ${\cal A}$ to be $\theta_{{\cal H}} \sim 0.19$ and $\theta_{{\cal A}} \sim 0.04$ arcsec, respectively, corresponding to physical radii of $R_{{\cal H}} \sim 701 \ \mathrm{pc}$ and $R_{{\cal A}} \sim 91 \ \mathrm{pc}$, at their respective preferred redshifts. Figure \ref{fig:data_and_snr} shows circles of this robust radius overlaid on zoomed-in maps of signal-to-noise ratio (SNR) of the two fields. These radii lie in relatively high-SNR regions around the perturber, where the data are most sensitive to the presence of low-mass haloes. This is particularly evident for detection ${\cal H}$, where the robust radius is large enough for this object to affect a higher-SNR part of the arc.

For a circularly symmetric surface mass density \( \Sigma(R) \propto R^{-\tilde{\gamma}(R)} \), we define the mass-weighted projected slope, analogous to the 3D definition in \cite{Dutton_2014}, as follows
\begin{equation} 
\label{eq:slope_1}
\gamma_{\text{2D}}  (R) = \frac{1}{M_{\text{2D}}(R)} \int_0^r \tilde{\gamma}(x) 2\pi x \Sigma(x) dx \,,
\end{equation}
which can be expressed in terms of the local logarithmic slope of the projected mass:
\begin{equation}
\label{eq:slope_2}
\gamma_{\text{2D}} (R) = 2 - \frac{2\pi R^2 \Sigma(R)}{M_{\text{2D}}(R)} = 2 - \frac{d \ln M_{\text{2D}}(R)}{d \ln R}. \end{equation}
For a singular isothermal profile, the projected slope is $\gamma_{\rm{2D}}(R) = 1$. For detection ${\cal H}$, the enclosed projected mass and slope at the robust radius $R_{{\cal H}}$ are $M_{\rm{2D}} = 3.44^{+0.33}_{-0.33} \times 10^9 M_\odot$ and $\gamma_{\rm{2D}} = 1.13^{+0.07}_{-0.07}$, respectively, based on the inferred perturber model with the highest Bayesian evidence. Similarly, for detection ${\cal A}$ at the robust radius of $R_{{\cal A}}$, the corresponding values are $M_{\rm{2D}} = 3.07^{+1.68}_{-1.48} \times 10^7 M_\odot$ and $\gamma_{\rm{2D}} = 0.84^{+0.13}_{-0.12}$. Note that the apparent weakness of the mass constraint here is purely a consequence of the distance uncertainty. The bending angle at the robust radius (which is proportional to $M_{\rm{2D}}/r_p$) is determined to better than 10 per cent. For all the models for which the evidence indicates a reasonable fit to the data, the mass and slope measurements remain stable within $\sim1\sigma$ at the robust scale (see Fig. \ref{fig:projected_profs}). 
By contrast, strongly disfavoured models for both detections show significant deviations in these quantities at this scale. 
In Sections \ref{sec:compare_with_cdm} and \ref{sec:compare_with_sidm}, we use $M_{\rm{2D}}$ and  $\gamma_{\rm{2D}}$ to compare the properties of both detections with expectations from different dark matter models.

\begin{figure}
    \centering
    \includegraphics[width=0.45\textwidth]{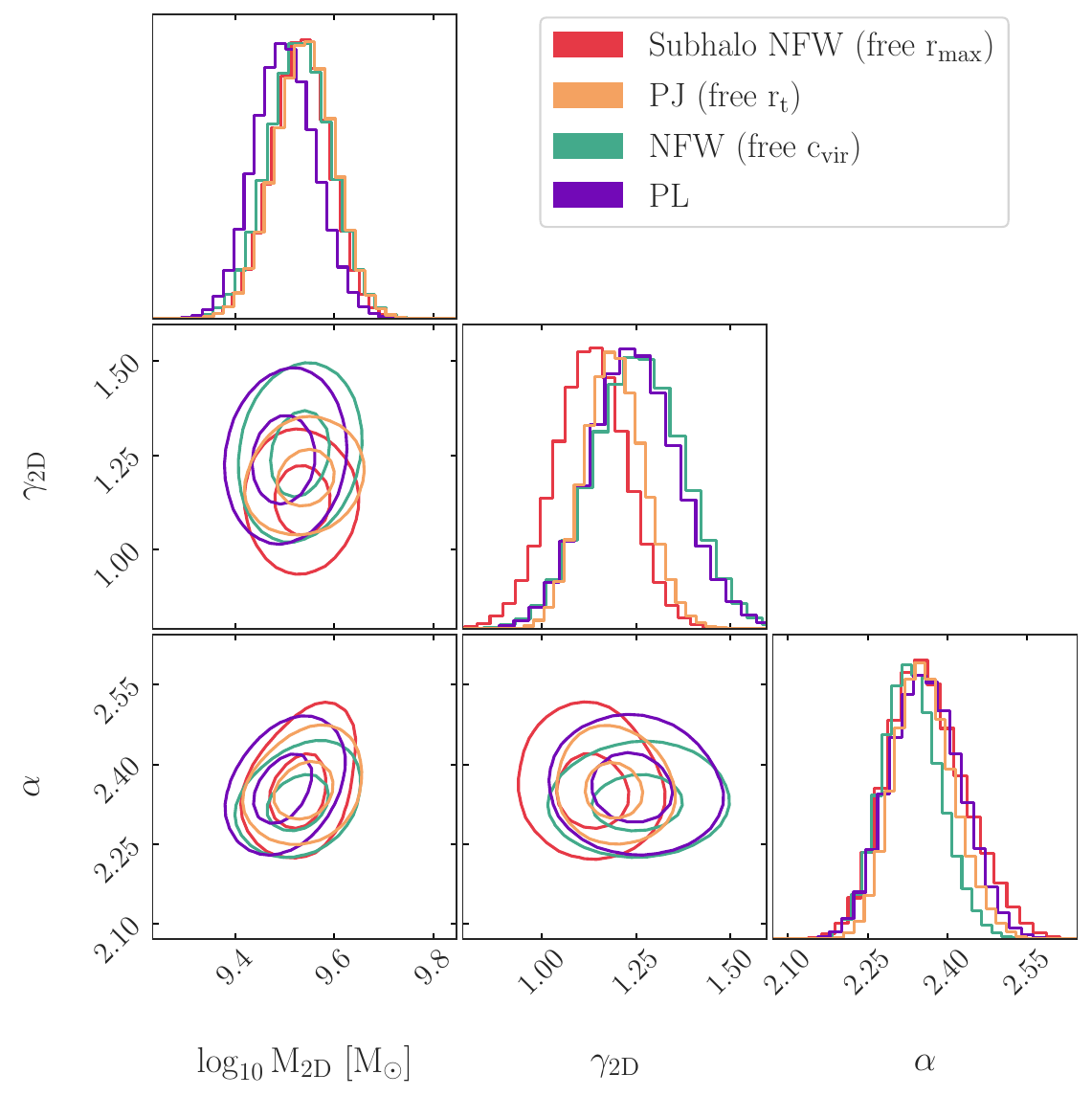}
    
    \vspace{1em} 

    \includegraphics[width=0.45\textwidth]{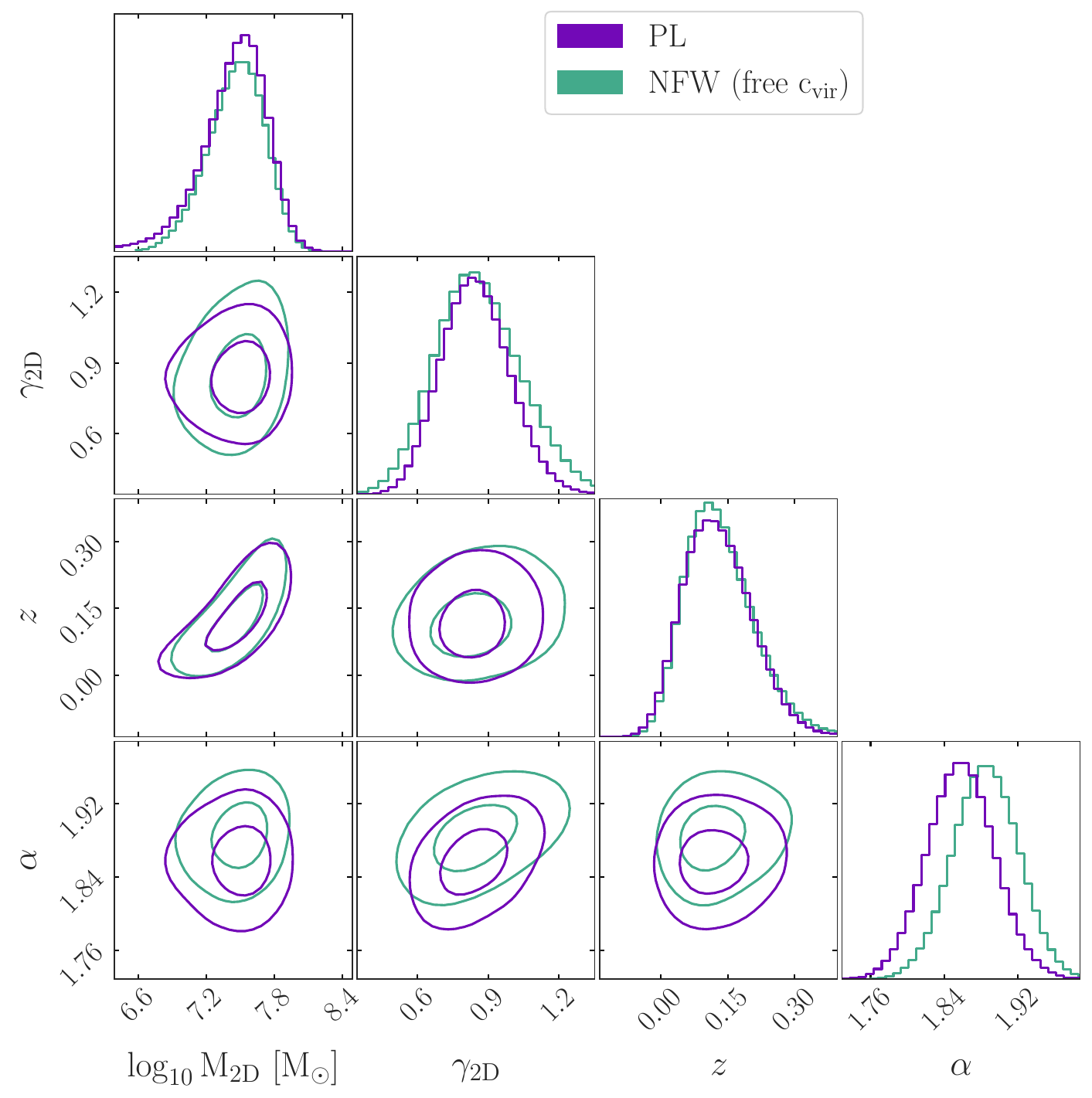}
    
    \caption{Projected properties for detection $\mathcal{H}$ (top) and $\mathcal{A}$ (bottom) and the main lens power-law slope $\alpha$.}
    \label{fig:2D_props_slope}
\end{figure}

We note that for models other than the PL, our definition of the slope differs from that used in previous works. Hence, a one-to-one comparison is not always possible. Similarly, our definition of robust radius is not the same as the one adopted by \cite{Minor_2021}, who defines the so-called perturbation radius as the distance between the centre of the detected object and the location of maximum perturbation on the critical curve. Their approach yielded a robust radius for detection ${\cal H}$ of $\sim 1$ kpc, slightly larger than the value used here. Our best-fitting models for this object yield $M_{\rm{2D}}(< 1~\rm{kpc}) \sim 3.7$–$5.1 \times 10^9 M_\odot$ at this radius, consistent with \cite{Minor_2021} and \citet{Despali_2024}. By modelling only the lowest-redshift source, \citet{Minor_2025} have found that larger values of the projected mass are allowed compared to the case with two sources. This may stem from a degeneracy between $M_{\rm{2D}}$ and the slope of the host lens $\alpha$ for this system (see Fig. \ref{fig:2D_props_slope}). 

For the PL model, we can directly compare our $\gamma_{\rm 2D}$ with \citet{Enzi_2024} and \citet{Minor_2025}; our inferred values are significantly shallower, as already found in other analyses that only model the lowest-redshift source. However, we also note from Fig. \ref{fig:2D_props_slope} that our slope is more tightly constrained (and not degenerate with $\alpha$) when compared to \citet{Minor_2025}. This is related to our different definition of robust radius and their choice of a truncated NFW for the mass density profile of this object. As the truncation radius is unconstrained, the prior-volume effect leads to overly broad posterior distributions.

\subsection{Constraints on the luminosities of the detected haloes}

Under the assumption that both detections are substructures of their central lens galaxies, \cite{Vegetti_2010} and \citet{Vegetti_2012} estimated $3\sigma$ upper limits on the luminosities of detections  ${\cal H}$ and ${\cal A}$ as $L_V < 5.0 \times 10^6 L_{V,\odot}$ and 
$L_V < 5.4 \times 10^7 L_{V,\odot}$, respectively. The luminosity of $\cal{H}$ was later revisited by \cite{Minor_2021}, who relaxed the upper bound to $L_V < 1.2 \times 10^8 L_{V,\odot}$ at 95 per cent confidence level, by directly modelling the subhalo's light profile. We adopt this more conservative upper limit on the luminosity of $\cal{H}$ throughout this paper.

Based on the luminosity distance corresponding to the average redshift of our best-fitting PL field halo model, we derive a new $3\sigma$ upper limit on the luminosity of detection $\mathcal{A}$ as $L_V < 6.3 \times 10^5 (z/0.13)^2L_{V,\odot}$, implying that this object is intrinsically nearly two orders of magnitude fainter than previously assumed.

\begin{figure}
\centering
\includegraphics[width=1.0\linewidth]{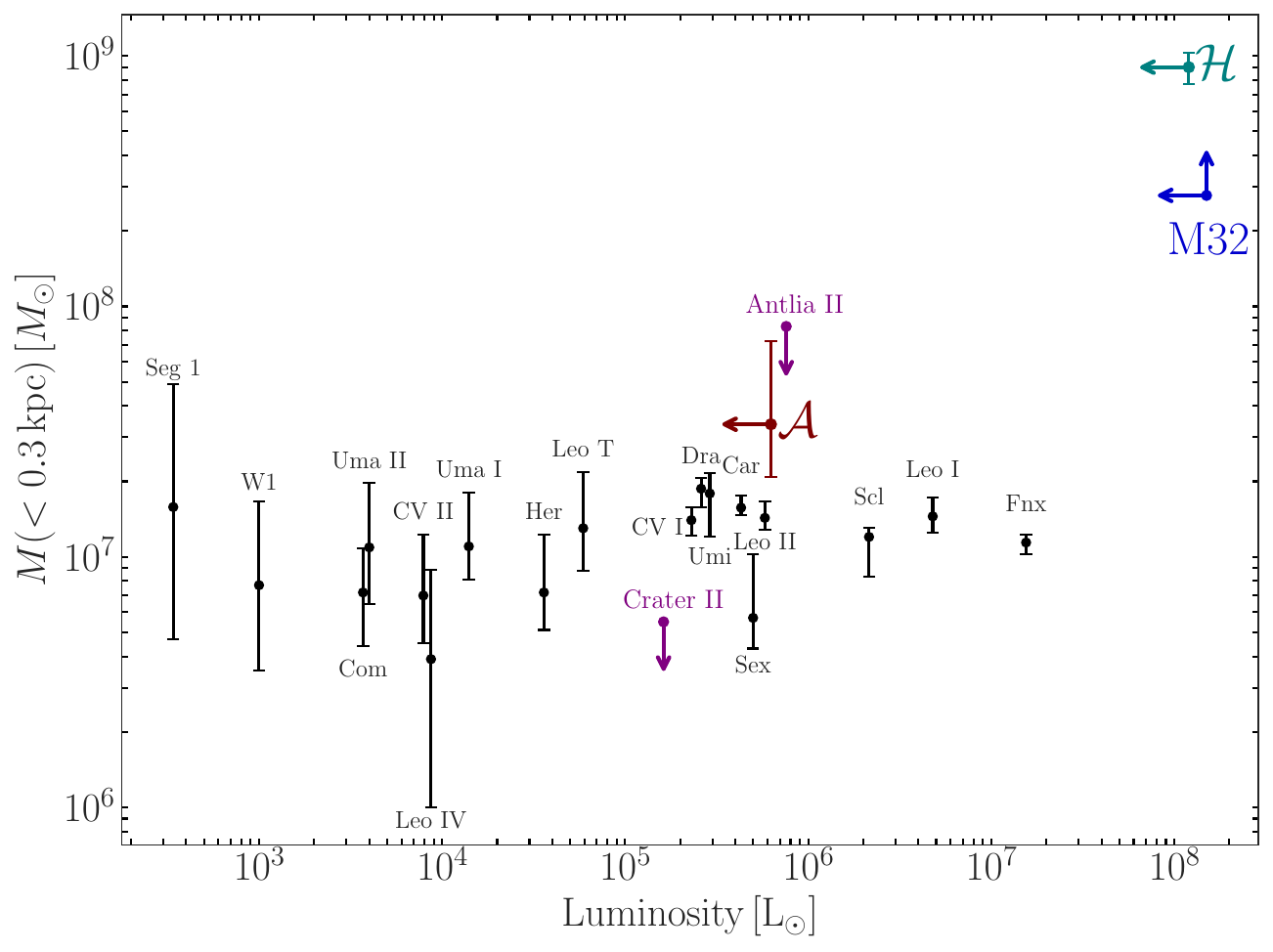}
\caption{Enclosed three-dimensional mass within 0.3 kpc as a function of total luminosity for detections ${\cal H}$ and ${\cal A}$, compared to the M32, and classical satellites and ultra-faint dwarf (UFD) galaxies of the Milky Way from \citet{Strigari_2008} and \citet{Ji_2021}. The mass and luminosity of M32 reflect its stellar mass and I-band luminosity \citep{Choi_2002, Verolme_2002}, and therefore provide a lower limit on its total mass and an upper limit on its V-band luminosity. In contrast, the masses of Antlia II and Crater II are dynamical estimates within their half-light radii, and thus represent upper limits on their masses enclosed within a 0.3\,kpc aperture.}
\label{fig:ML_ratio}
\end{figure}

Fig. \ref{fig:ML_ratio} compares the mass and luminosity of M31's subhalo, M32, and of the Milky Way's dwarf satellites with those of detections ${\cal H}$ and ${\cal A}$, showing that the latter lies in a region of the $M$–$L$ plane consistent with the Galactic satellite population. Instead, detection ${\cal H}$ is almost two orders of magnitude more massive than these satellites, in agreement with the findings of \cite{Vegetti_2010}, but comparable to M32, which is by far the most concentrated dwarf galaxy in the local Universe. It is worth noting that detection ${\cal H}$ has a circular velocity of $\sim 123$ km/s at $r \sim 750$ pc that is significantly larger than those measured for M33 and the Large Magellanic Cloud at similar radii, while the upper limit on its luminosity is smaller by about an order of magnitude.
    
\subsection{Macro-models}
\label{sec:macro}

Here, we discuss the properties of the main deflector for our overall best-fitting models of detections ${\cal H}$ (an NFW subhalo) and ${\cal A}$ (a PL field halo). 

We find that the data for J0946+1006, strongly favour the presence of angular complexity in the main lens, with increases in the log-evidence of $\Delta\log{\cal{E}}_{3,4} = 22.3$ and $\Delta\log{\cal{E}}_{1,3,4} = 33.9$ upon including multipoles of orders $m=3, 4$ and $m=1, 3$ and 4, respectively. The median multipole amplitudes, defined as $\eta_m = \sqrt{a_m^2 + b_m^2}$, for $m=3$ and $m=4$ are both at the level of 1 per cent, with $\eta_3 = 0.010^{+0.002}_{-0.001}$ and $\eta_4 = 0.011^{+0.002}_{-0.002}$, while the $m = 1$ multipole has a higher amplitude of $\eta_1 = 0.093^{+0.006}_{-0.011}$. Our results are consistent within $1-2 \sigma$ with those obtained by \citet{Enzi_2024} when modelling the lowest-redshift source only, as we do here. Unlike, \citet{Minor_2025} we find that the inclusion of multipoles leads to improved image residuals and that the increase in Bayesian evidence is not solely related to a smoother source.

Different analyses of this system \citep{Gavazzi_2008, Vegetti_2010, Sonnenfeld_2012, Collett_2014, Minor_2021, Etherington_2023, Turner_2024, Ballard_2024, Despali_2024, Enzi_2024, Minor_2025} infer values for the power-law density slope that are inconsistent given their reported uncertainties, ranging from $\alpha = 1.95\pm0.02$  \citep{Sonnenfeld_2012} to $\alpha = 2.58\pm 0.02$ \citep{Despali_2024}. Our own analysis finds $\alpha = 2.35^{+0.08}_{-0.04}$ from our best-fitting model with the NFW subhalo.

These analyses employ a number of different methods to measure the power-law slope which explains their discrepant results. Measurements that combine lensing information with stellar dynamics can measure the slope reliably over a large radius, but with comparatively larger uncertainties than a lensing-only analysis. It can be inferred from lensing information alone if an extended source model is used \citep[e.g.][]{Vegetti_2010,Despali_2024,Enzi_2024}, essentially due to the radial extension of the ring or rings, but this measurement is vulnerable to the mass-sheet transformation \citep[MST, e.g.][]{Schneider_2013}. In this case, the slope is only being measured close to or on the ring and extrapolated to the rest of the profile using the model chosen to break the MST, especially if the lens has a low ellipticity \citep{ORiordan2021}. In this type of analysis, it should be no surprise that including or excluding rings at different radii from the modelling can lead to very different inferences of the slope. The MST also means that choices made in source reconstruction will affect the inferred slope. For example, a reconstruction which allows a smaller (larger) source can reconstruct the same imaging data by changing the local magnification with a shallower (steeper) slope. Differences in angular mass complexity, lens light modelling, and underestimated uncertainties also contribute to the disagreements in the reported values. We refer the interested reader to \cite{Minor_2025} for further discussion on the matter for the specific case of J0946+1006.

For the lens system B1938+666, we find that including mass multipoles of order $m=3,4$ leads to an increase in the Bayesian evidence of $\Delta\log{\cal{E}}_{3,4} = 5.7$. In contrast, the data strongly disfavour a multipole of order $m=1$. The median amplitudes of the $m = 3$ and $m = 4$ multipoles in our best reconstruction are $\eta_3 = 0.002^{+0.001}_{-0.001}$ and $\eta_4 = 0.004^{+0.001}_{-0.001}$, respectively. Our findings highlight the diversity in the angular structure of lens galaxy mass distributions, with different lenses displaying different levels and types of angular complexity, likely related to their formation histories. 

From our best-fitting model (i.e. with $\cal{A}$ as a PL field halo), we find a power-law slope of $\alpha = 1.86^{+0.03}_{-0.03}$ for the main lens. This is consistent with \citet[][$\alpha = 1.89 \pm 0.02$]{Despali_2024}, and significantly shallower than the values inferred by \citet[][$\alpha = 2.058\pm0.013$]{Vegetti_2012} and \citet[][$\alpha = 2.322$]{Seng_l_2022}.

\section{Comparison with CDM Predictions}
\label{sec:compare_with_cdm}

According to CDM $N$-body simulations, dark matter haloes are generally well approximated by an NFW mass density profile \citep[e.g.][]{Springel_2008} with a fairly well-defined relation between their mass and concentration \citep[e.g.][]{Duffy_2008, Molin__2023}.

\begin{figure}
    \centering
    \includegraphics[width=1.0\linewidth]{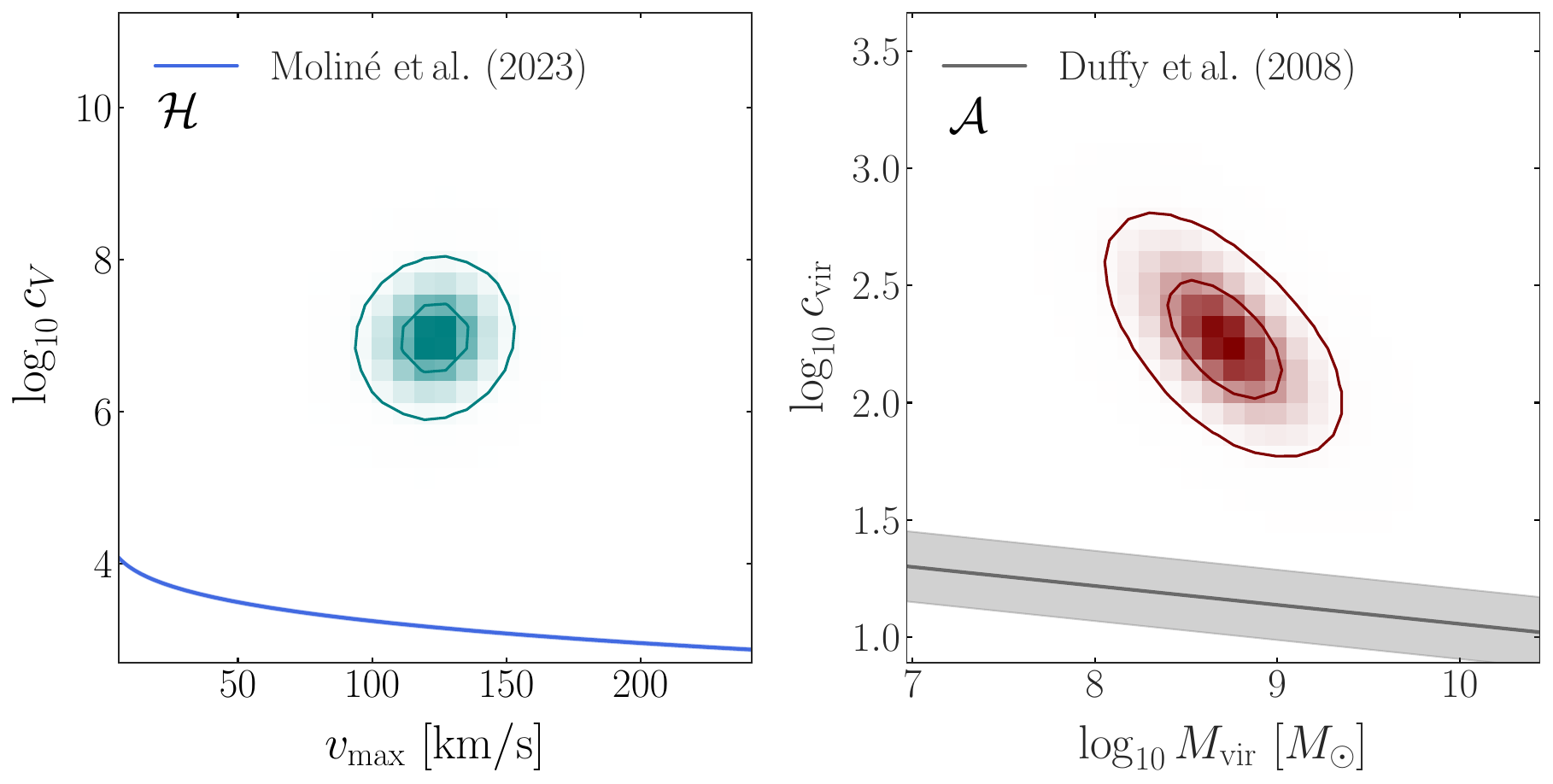}
    \caption{Comparison of the NFW parameters for both detections with predictions from $N$-body simulations. Left: posterior distribution of $\vmax$ and $c_V$ for detection $\mathcal{H}$, together with the corresponding $\vmax$–$c_V$ relation for subhaloes from \citet{Molin__2023} (blue curve and shaded $1\sigma$ band). Right: posterior distribution of $\mvir$ and $\cvir$ for detection $\mathcal{A}$, alongside the mass–concentration relation from \citet{Duffy_2008} at the perturber’s inferred redshift of $z = 0.12$ (grey curve and shaded $1\sigma$ band).}   
    \label{fig:concentration_comparison}
\end{figure}

Fig. \ref{fig:concentration_comparison} shows the posterior distributions of the NFW fit parameters for both detections, together with the mass–concentration (or equivalently, $\vmax-\rmax$) relations derived from CDM $N$-body simulations. Consistent with previous studies, we find that the inferred concentration of detection ${\cal H}$ is inconsistent with CDM dark matter-only predictions. 
For a subhalo at $z=0.22$ with a peak circular velocity of $\vmax = 123$ km/s, the expected concentration for subhaloes is $\log_{10} c_\mathrm{V} = 3.75 \pm 0.02$ \citep{Molin__2023}. However, the inferred concentration from the data is $\log_{10} c_\mathrm{V} = 6.97^{+0.21}_{-0.22}$, corresponding to a deviation of $\sim 15\sigma$ from the predicted median. Here, $\sigma$ is the quadrature sum of the uncertainty on the inferred value and the scatter in the lognormal relation from \citet{Molin__2023}. 

The expected concentration of a CDM halo with mass and redshift consistent with detection ${\cal A}$, i.e. $\mvir = 5.08^{+3.86}_{-2.20} \times 10^{8} M_\odot$ and $z=0.12$, is $\log_{10} \cvir = 1.16 \pm 0.15$ \citep{Duffy_2008}. The median concentration inferred from our analysis is $\log_{10}\cvir = 2.27^{+0.21}_{-0.19}$, corresponding to a $\sim 4\sigma$ deviation from the CDM model, where $\sigma$ is the combined scatter from the measurement and the lognormal distribution given by \cite{Duffy_2008}.

Hence, while the discrepancy is less extreme for detection ${\cal A}$ than for detection ${\cal H}$, both objects remain in significant tension with expectations from dark-matter-only (DMO) $\Lambda$CDM $N$-body simulations.

Although both detections are dark matter-dominated, they are both relatively massive and, therefore, likely host some baryons. Hence, it is more appropriate to compare their properties with predictions from hydrodynamical simulations, where baryons can significantly alter the profile of dark matter haloes. Moreover, as we demonstrated in Section \ref{sec:modelling_results}, the NFW parameters are not the quantities that are most robustly inferred and it is better to focus instead on the projected mass density profile around the robust radius.

In the rest of this section, we compare the properties of detections ${\cal H}$ and ${\cal A}$ with those of haloes in the hydrodynamical simulation TNG50-1 \citep{Nelson_2019, Pillepich_2019}.  

\subsection{Analogues of detection \texorpdfstring{$\mathcal{H}$}{H} in TNG50}

\label{sub:analogues_h}

We draw the comparison sample for detection $\mathcal{H}$ from subhaloes in the TNG50-1 simulation, extracted using the \textsc{subfind} algorithm \citep{Springel_2001} at $z \sim 0.22$ (snapshot 82). We select systems hosted by haloes with \textsc{subfind} masses in the range $10^{12} \leq M_{\rm{tot}}/M_\odot \leq 5 \times 10^{13}$ and stellar masses between $10^{10} \leq M_\star/M_\odot \leq 10^{13}$. These ranges are motivated by the estimated mass of the main deflector in the J0946+1006 lens system, as reported by \citet{Auger_2010}, who find a velocity dispersion within half of the effective radius of $\sigma_{e/2}\sim265$ km/s and a total stellar mass of $\log (M_\star/M_\odot)\sim11.2-11.7$. 
From this subset, we further limit our sample to subhaloes with $5 \times 10^9 < M_{\rm{tot}}/M_\odot < 5 \times 10^{12}$ and a minimum of 500 dark matter particles. 
Our selection of host galaxies encompasses the samples of \citet{Minor_2021} and \citet{Despali_2024}, but reaches lower total masses. Our selected subhaloes extend an order of magnitude higher in mass, ensuring coverage of the virial mass inferred from our fixed-concentration NFW model.

As shown in Section \ref{sec:modelling_results}, the properties of detections that are most robustly constrained are the projected mass and slope within a range of radii around their robust scale. Both quantities are consistent within the errors among all mass density profiles that give an acceptable fit. Additionally, the relative error on the deflection angle reaches its minimum at this scale. Hence, we classify as analogues of detection $\cal{H}$ those simulated subhaloes with projected mass and slope within the ranges $1.5\times10^9 \leq M_{\rm{2D}}(\theta_{\cal{H}}=0.19\arcsec)/M_\odot\leq 8\times10^{9}$ and $0.9 \leq \gamma_{\rm{2D}}(\theta_{\cal{H}}=0.19\arcsec) \leq1.5$, where $\theta_{\cal{H}}$ corresponds to a comoving distance of $r_{\rm{com}} = 871$ pc at the simulation snapshot considered. 

The projected mass and the local logarithmic slope used to compute $\gamma_{\rm 2D}$ for each halo are measured within a cylindrical aperture of radius $\theta_{\cal{H}}$, with the slope estimated via a finite difference between neighbouring radii. Both quantities are averaged over 1000 random lines of sight. 

\subsection{Analogues of detection \texorpdfstring{$\mathcal{A}$}{A} in TNG50}

\label{sub:analogues_a}

We first select from the TNG50-1 simulation central galaxies within the redshift range $z=0.06$ to $z=0.35$, corresponding to simulation snapshots 75 to 94. These values encompass the lower $1 \sigma$ and upper $3 \sigma$ limits of the inferred redshift. These galaxies are identified by \textsc{subfind}, in which the central or \emph{primary} halo of each FOF group is, by assumption, the most massive among the members of the group. We then restrict our selection to central galaxies within the \textsc{subfind} mass bin $2\times{10}^8 <M_{\rm{tot}}/M_\odot<5\times{10}^{11}$ and require that haloes contain at least 500 dark matter particles. We note that no central galaxy with total mass below $2\times10^8 M_\odot$ meets the latter threshold; we exclude such low-mass objects from our analysis to avoid excessive particle noise.

Analogues of detection $\cal{A}$ are defined as simulated haloes with a projected mass and slope within the $1\sigma$ scatter of the values inferred in Section \ref{sec:modelling_results}, that is, $1.5 \times 10^7 < M_{\rm{2D}}(\theta_{\cal A}=0.04\arcsec)/M_\odot <5.0 \times 10^7$ and $0.7 < \gamma_{\rm{2D}}(\theta_{\cal A}=0.04\arcsec) < 1.05$.

For the selected redshift range in the simulation, the robust radius of  0.04 arcsec corresponds to a comoving radius of $r_{\rm{com}} = 49.5 - 273.5$ pc, and lies, therefore, below the scale at which one can reliably measure the mass profile of simulated haloes in TNG50-1 (commonly estimated as $\sim 2.8$ times the Plummer equivalent gravitational softening length of the dark matter particles, where $\epsilon_{\rm{DM},\star} = 288$ pc at $z=0$). For this reason, we first need to fit analytical profiles to the mass density profiles of selected objects and extrapolate them down to the regions of interest. In Appendix \ref{sec:profile_fitting}, we provide more details on this procedure.
Briefly, our results show that the measured density profiles of the majority of central haloes in our sample are best fit by the Einasto profile, in the sense that it yields the lowest {\it rms} residual among the models considered, with the generalized NFW profile (hereafter gNFW) being the next favoured model. This indicates that most simulated haloes are more accurately described by a smooth, gradual slope transition between the inner and outer regions, rather than a sharp break. We find a median Einasto shape parameter of $\alpha_{\rm e} \sim 0.22$ for haloes in the sample. Relating the shape parameter of the Einasto profile to halo concentration requires careful consideration, as concentration is best characterised in terms of the peak of the circular velocity \citep[e.g.,][]{Prada_2012}. In general, for an Einasto profile, the concentration depends both on the shape parameter and on the characteristic radius of the halo \citep{Klypin_2016}. 

For each halo, we adopt the profile yielding the lowest fit rms and calculate the projected enclosed mass and slope at $\theta_{\cal A}$ from its analytical form. 

\subsection{Comparing analogues to observations}

\begin{figure*}
    \centering

    \includegraphics[width=0.495\textwidth]{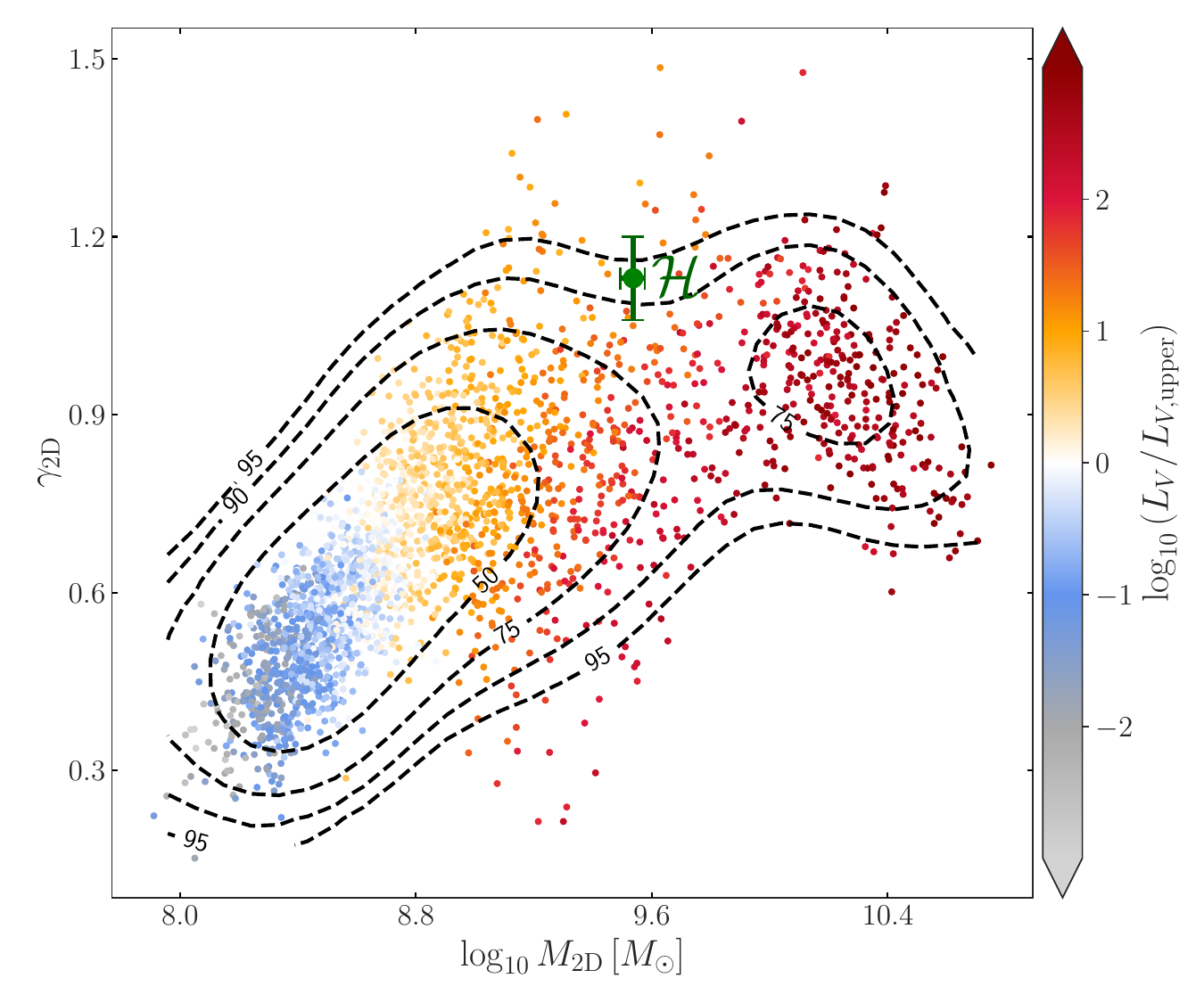}
    \hfill
    \includegraphics[width=0.495\textwidth]{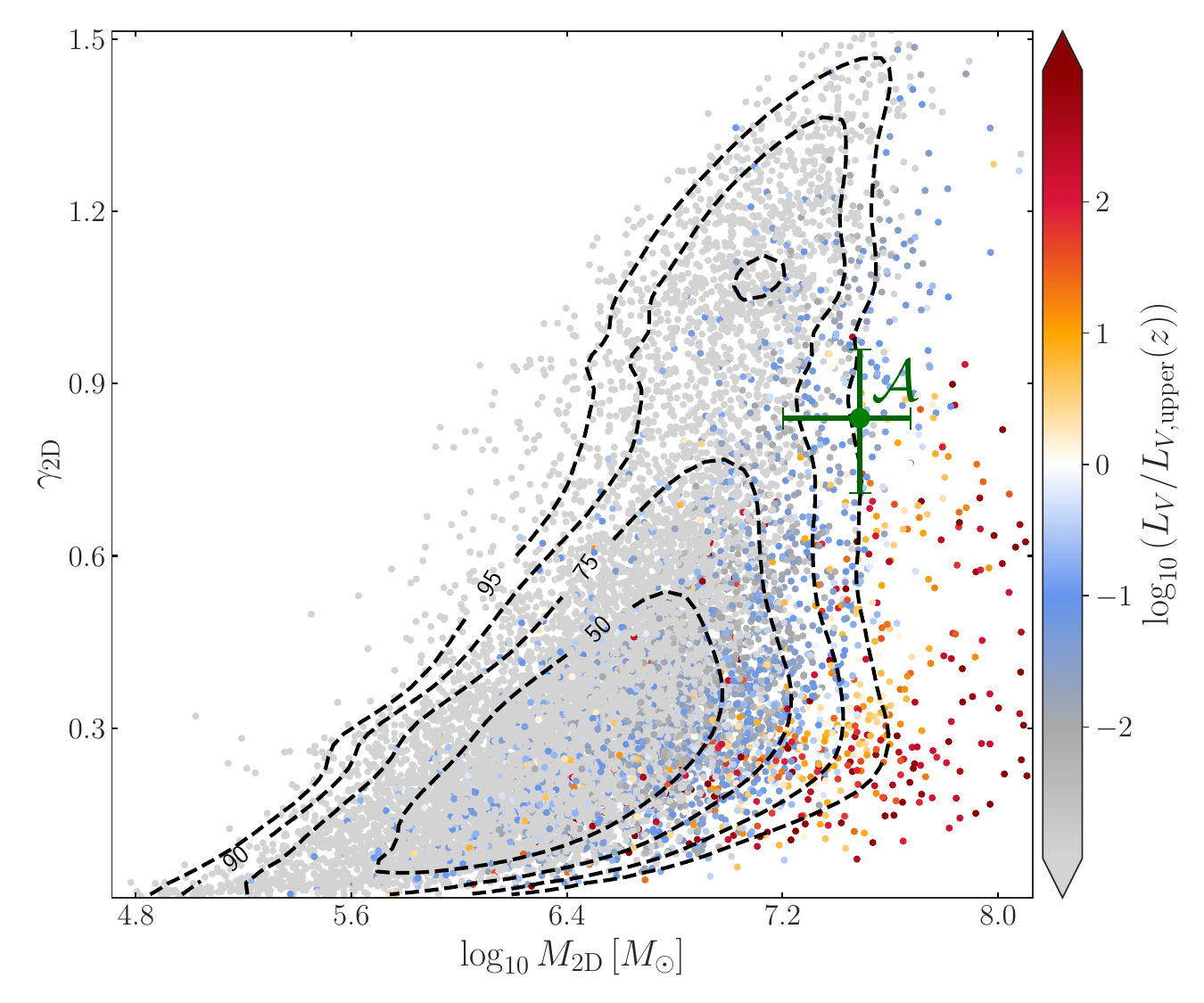}

    \caption{Left: Projected enclosed mass and slope at the robust radius for all subhaloes in the TNG50-1 simulation at redshift $z \sim 0.22$ and total mass $5\times{10}^9\leq M_{\rm{tot}}/M_\odot\leq 5\times{10}^{12}$. Their hosts have total masses in the range ${10}^{12}\leq M_{\rm{tot}}/M_\odot\leq 5\times{10}^{13}$ and central galaxy stellar masses between ${10}^{10}\leq M_{\star}/M_\odot\leq {10}^{13}$. The dashed contours enclose 50, 75, 90, and 95 per cent of the cumulative probability distribution. The green point represents detection $\cal{H}$. Points are colour-coded by the ratio $\log_{10} \,(L_V/L_{V, \rm{upper}})$, where the upper luminosity bound for detection $\cal{H}$ is $L_{V, \rm{upper}} = 1.2 \times 10^8 L_{V,\odot}$.
    Right: Same as the left panel, but for a random 0.2 per cent sample of central haloes in the TNG50-1 simulation with redshift between $z=0.06$ and $z=0.35$ and total mass $2\times{10}^8\leq M_{\rm{tot}}/M_\odot\leq 5\times{10}^{11}$. Density contours are based on the full sample. The green point represents detection $\cal{A}$. The luminosity upper limit at each simulation redshift, $L_{V, \rm{upper}}(z)$, is derived from $L_{V,\rm upper} = 5.4 \times 10^7 L_{V,\odot}$ at $z = z_{\rm lens}$, adjusted using the corresponding luminosity distance.}
    \label{fig:2d_density}
\end{figure*}

Fig. \ref{fig:2d_density} shows the two-dimensional distribution of the projected enclosed mass and slope at the robust radius for the selected populations of simulated subhaloes and central haloes. Note that these correspond to the full selected samples, not just the samples of analogues.
Both detections fall at the edge of their respective 2D distributions in the relevant mass and redshift ranges, with detection $\cal{H}$ located outside the 93 per cent enclosed region (corresponding to the outermost $\sim7$ per cent) and detection $\mathcal{A}$ outside the 96 per cent enclosed region (outermost $\sim4$ per cent).  Their density profiles are therefore consistent with CDM but not representative of the typical (sub)halo populations in the simulation. Notably, the more massive haloes in both samples, whose projected properties are close to those of the detected objects in the two-dimensional parameter space, have shallower slopes at their robust radii compared to the less massive ones.
 
Based on the analogue criteria defined earlier, which require both the projected enclosed mass and slope at the robust radius to fall within specified ranges, a total of 126 simulated subhaloes have a projected mass density profile at the robust radius consistent with that inferred from the data for detection $\cal{H}$. This number increases to 365 when matching only the projected enclosed mass to the observed value. 
For detection $\cal{A}$, approximately $1.4\times 10^5$ central haloes meet both conditions, rising to $6.5\times 10^5$ when considering mass alone. The median values of the gNFW slope and Einasto shape parameter for these haloes are $\beta = 1.89$ and $\alpha_{\rm e} = 0.11$, deviating significantly from the inner slope of an NFW profile ($\beta = 1$) and consistent with the steeper profile inferred from the observation.

\begin{figure*}
	\includegraphics[width=1.0\textwidth]{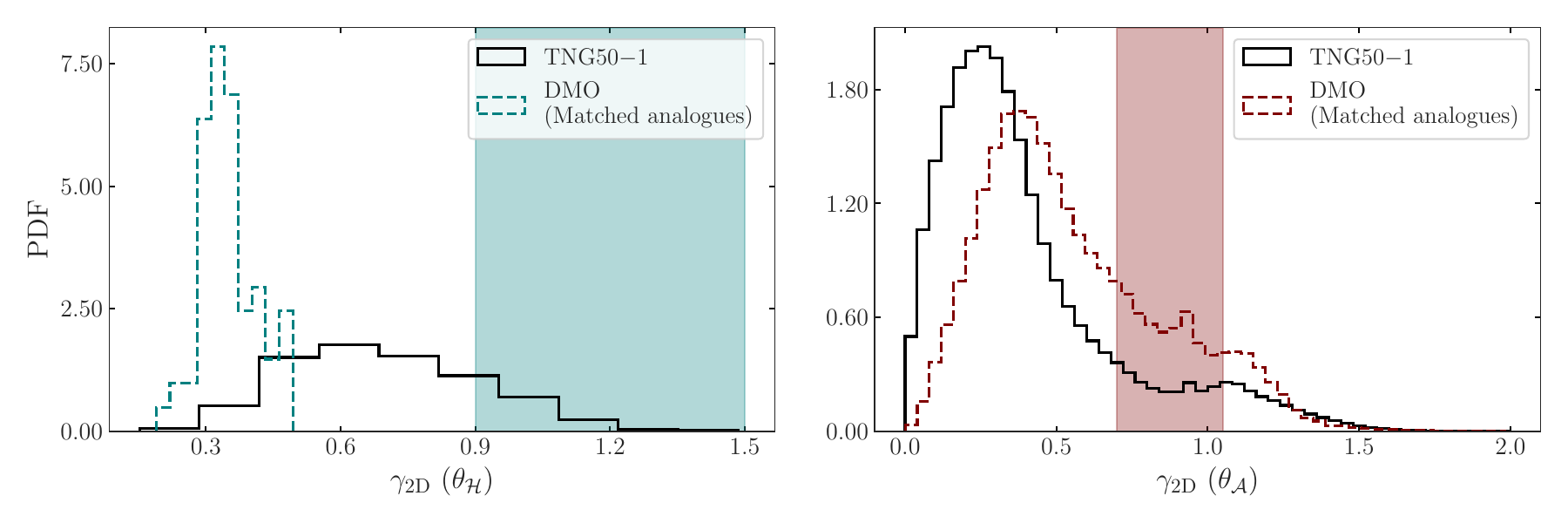}
    \caption{Distribution of the projected slope at the robust radius for the full sample of simulated haloes in the TNG50-1 hydro run (solid black histograms) for detections ${\cal H}$ (left) and ${\cal A}$ (right). The dashed histograms show the slope distribution for the dark-matter-only counterparts of analogues of each detection. The shaded bands indicate the range of slopes consistent with the inferred values for each object.}
    \label{fig:dist_2D}
\end{figure*}

The probability density functions of $\gamma_{\rm{2D}}$ at the robust radius for the two simulated halo samples are shown in Fig. \ref{fig:dist_2D}. The selected halo samples have median projected slopes of $\gamma_{\rm{2D}}(\theta_{\cal{H}}) = 0.68$ and $\gamma_{\rm{2D}}(\theta_{\cal{A}}) = 0.32$, respectively, both values are shallower than the slopes inferred for the observed perturbers. We note that the steeper slope inferred by \cite{Enzi_2024} and \cite{Minor_2025} would place object $\cal{H}$ further out in the tail of the distribution. However, because of the prior-volume effect related to their choice of mass density profile their posterior distribution may be overly broad and their confidence intervals on $\gamma_{\rm 2D}$ could be biased.

To evaluate the impact of baryons on the halo mass distributions, we now match each analogue halo in the hydrodynamic run to its counterpart in the corresponding dark-matter-only simulation. We do so by using the \textsc{LHALOTREE} algorithm as described in \cite{Nelson_2015}, which tracks unique dark matter particle IDs and matches haloes to their DMO variants based on the highest fraction of shared particles. However, not every halo has a well defined match, and some central haloes in the hydrodynamic run are no longer classified as such in the DMO run. Overall, we successfully matched 53 per cent of the analogues of $\cal{H}$ and 88 per cent of the analogues of $\cal{A}$ to their counterparts in the DMO run. 
Fig. \ref{fig:dist_2D} shows the distributions of $\gamma_{\rm{2D}}$ at the robust radius for the DMO variants of the detection analogues.
None of the DMO counterparts of the analogues of detection $\mathcal{H}$ reproduces its density profile, whereas 14.5 per cent of those for $\cal{A}$ remain consistent with the observed halo. The projected slopes of the haloes in the DMO run whose baryonic counterparts have density profiles consistent with the inferred ranges for the detections (shaded regions in Fig. \ref{fig:dist_2D}) are now clearly skewed towards lower values, with median slopes of $\gamma_{\rm{2D}}(\theta_{\cal{H}}) = 0.34$ and $\gamma_{\rm{2D}}(\theta_{\cal{A}}) = 0.49$. This result is consistent with previous studies \cite[e.g.,][]{Schaller_2015} showing that the inclusion of baryons can lead to steeper inner density profiles due to contraction.

\begin{figure}
    \centering
    \includegraphics[width=\linewidth]{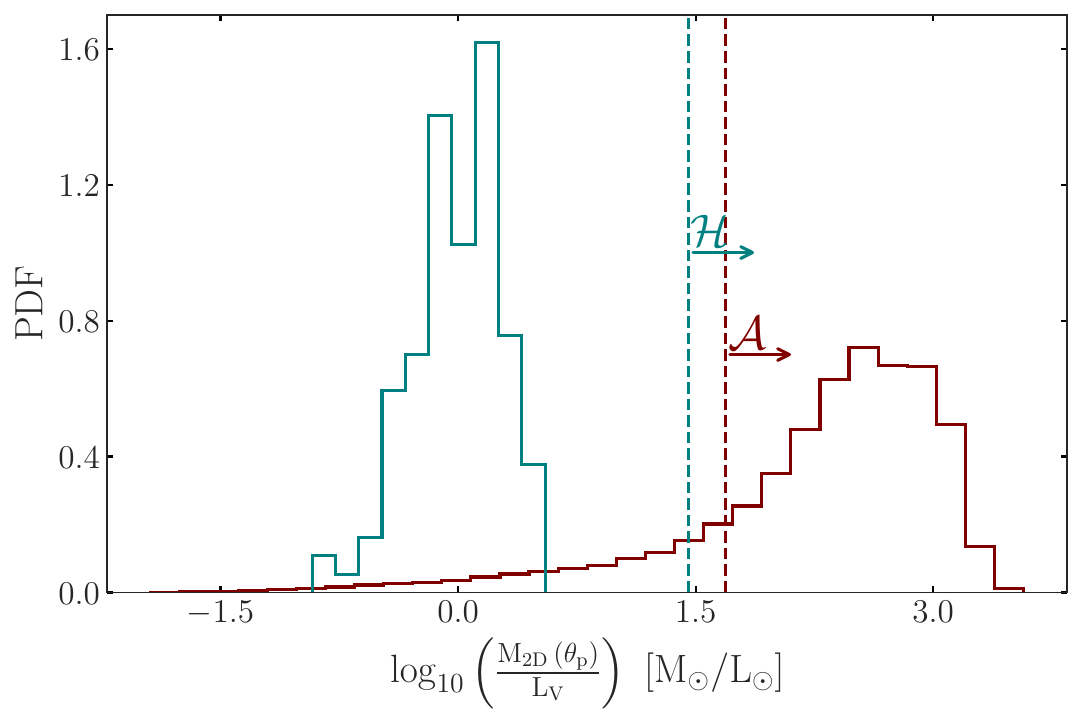}
    \caption{Distribution of the mass-to-light ratio for simulated (sub)haloes whose projected masses and mass density profiles are consistent with objects $\cal{H}$ (teal histogram) and $\cal{A}$ (red histogram). The vertical dashed lines indicate the lower limits on the mass-to-light ratio for each object, shown in the corresponding histogram colour.}
    \label{fig:ML}
\end{figure}

Up to this point, we have compared our observations with simulated haloes in terms of their mass density profile. We now also take into account the upper limits on the luminosity of the detected objects. We calculate the $V$-band luminosities of the simulated haloes using the photometric magnitudes provided in TNG50-1. Since no dust attenuation is applied in computing these photometric parameters, the resulting luminosities represent intrinsic, dust-free values. None of the analogue haloes satisfy the luminosity constraint for detection $\cal{H}$, even when adopting the conservative upper bound from \citet{Minor_2021}. This is consistent with the findings of \citet{Minor_2021} and \citet{Despali_2024}. Fig. \ref{fig:ML} shows the distribution of mass-to-light ratio for analogues of the detections, calculated based on projected mass within their robust radii. Unlike $\cal{H}$, most analogues satisfy the upper luminosity limit set by the observation, with only 8 per cent exceeding it and thus being ruled out. In particular, no haloes with $M_{\rm tot} > 2.2\times10^{10} M_\odot$ satisfy the constraint. As a result, applying a luminosity cut slightly reduces the number of TNG50-1 haloes matching the properties of detection ${\cal A}$ to approximately $1.3\times 10^5$, or to $\sim5.2\times 10^5$ when considering only haloes whose projected mass at the robust radius and luminosity are consistent with those inferred for the detection.

\section{COMPARISON WITH SIDM PREDICTIONS}
\label{sec:compare_with_sidm}

\begin{figure*}  
    \centering
    \begin{subfigure}[t]{0.49\textwidth}
        \centering
        \includegraphics[width=1\textwidth]{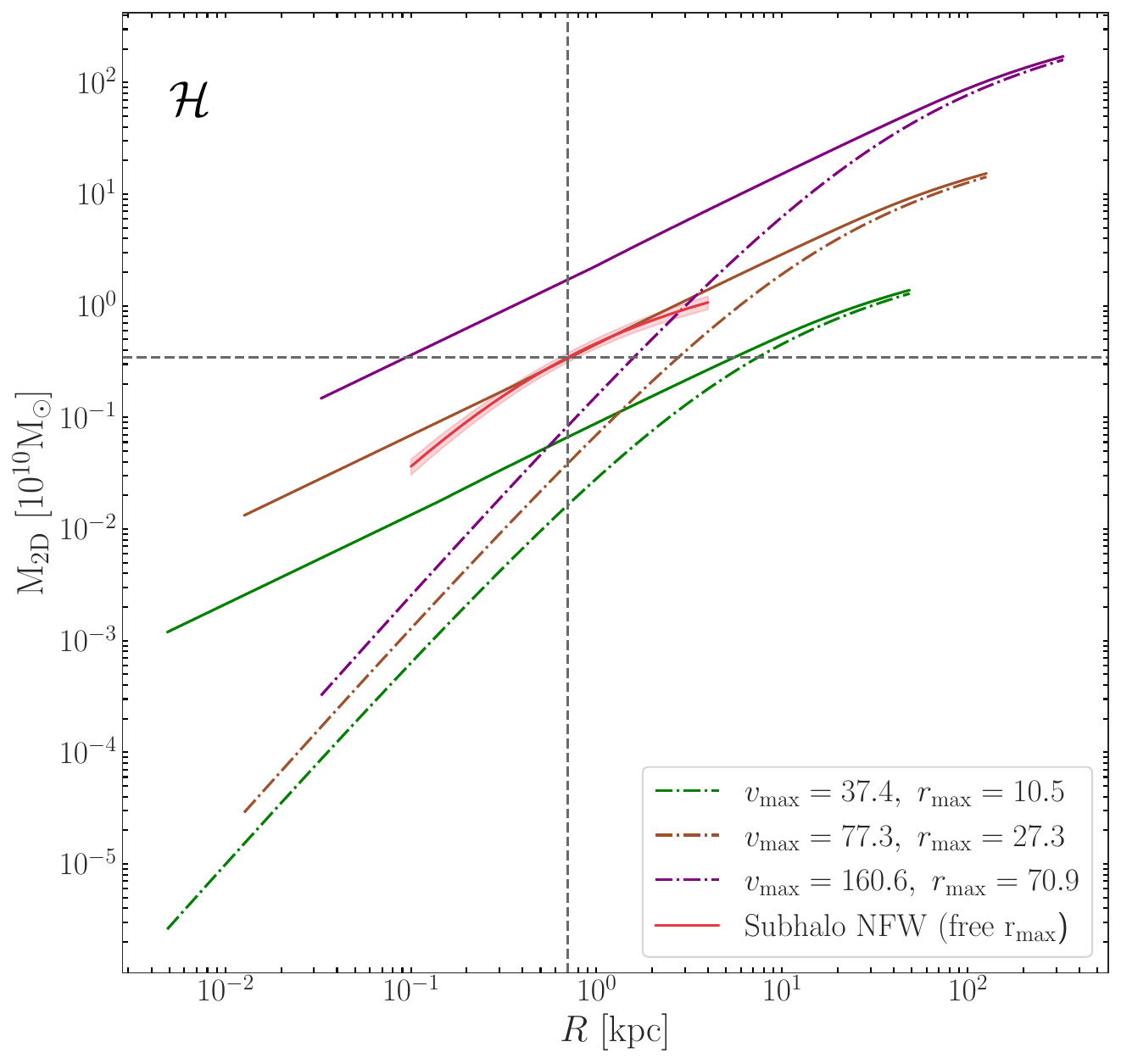}
        \label{fig:sidm_J0946}
    \end{subfigure}%
    \hfill
    \hspace{0.01\textwidth}
    \begin{subfigure}[t]{0.49\textwidth}
        \centering
        \includegraphics[width=1\textwidth]{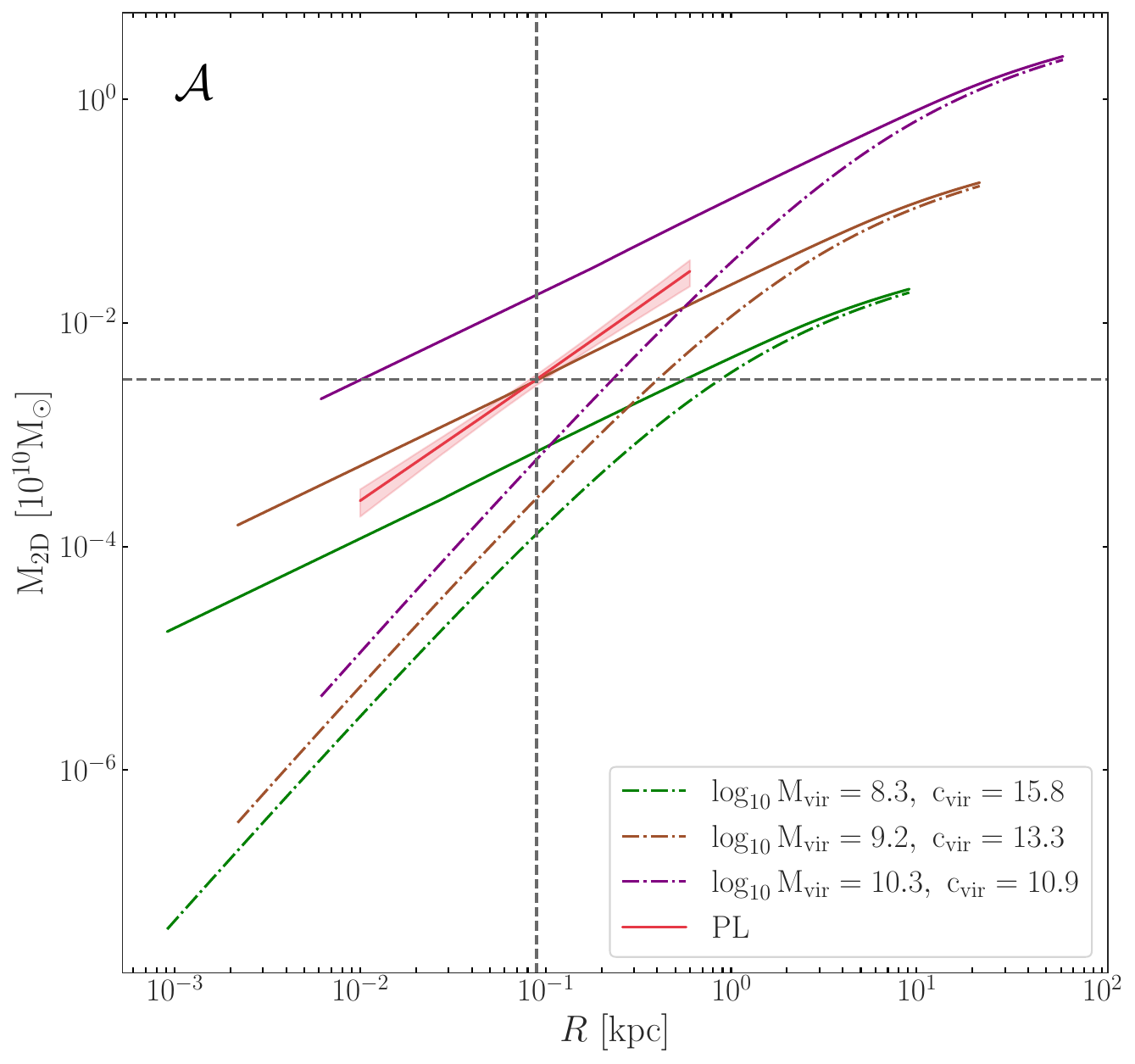}
        \label{fig:sidm_b1938}
    \end{subfigure}

    \caption{Projected mass profiles for different SIDM haloes (solid lines), for their CDM NFW analogues (dashed-dotted lines) and our best-fit model for detection ${\cal H}$ (left) and detection ${\cal A}$ (right). The vertical and horizontal lines represent the radius at $z=0.12$ and
    $z=0.222$ at which these models are, respectively, most 
    precisely constrained, and their corresponding projected 
    masses. The NFW profiles are labelled according to their 
    log values of $\vmax$ and $\rmax$ for the subhalo, and $\mvir$ and $\cvir$ for the field halo.}
    \label{fig:sidm_comparison}
\end{figure*}

It has been shown that the presence of non-gravitational interactions between dark matter particles initially leads to heat transfer from the hotter outer regions of a halo to its cooler centre, lowering the central density and forming a small core \citep[][]{Vogelsberger_2012, Rocha_2013, Zavala_2013}. This core expands and heats until the inner halo reaches a quasi-isothermal state, where the halo spends most of its pre-collapse lifetime \citep{Outmezguine_2023}. Once the temperature gradient is negative everywhere, the core begins to transfer heat outwards and to contract again, becoming increasingly dense in a rapidly accelerating, runaway process known as gravothermal collapse. In the later phases of this collapse, densities in the inner regions significantly exceed their initial NFW values, but these phases are short and culminate in the formation of a black hole which then grows rapidly through Bondi accretion until it has swallowed the entire inner part of the halo where particle mean free paths are short and a fluid approximation is valid \citep{Pollack_2015}.\footnote{In this context it is interesting that M32, by far the densest known dwarf galaxy in the local Universe, and hence, perhaps, the most likely site for complete SIDM core collapse, contains a central black hole of mass $2.5\times 10^6M_\odot$, almost 1 per cent of its stellar mass \citep{Verolme_2002}.} As a result, dark matter haloes in self-interacting dark matter (SIDM) models display a large variety of mass profiles, depending on the stage of their evolutionary path at which they are observed.

In this section, we compare the observed properties of detections ${\cal H}$ and ${\cal A}$ with theoretical predictions for self-interacting dark matter. Specifically, we focus on the theoretical framework based on fluid models developed by \cite{Outmezguine_2023} for the (nearly) universal gravothermal evolution of an idealised and isolated SIDM halo. Using a fit to the scaled density profile of a SIDM halo in the late stages of its evolution, specifically at time $t \simeq 388 t_{c,0}$ for their $n = 3.7$  model, where $t_{c,0}$ is the characteristic time scale, we derive the late-time projected mass profile of SIDM haloes that were initially described by an NFW profile. We then compare this with the mass profiles inferred for our detected objects across a range of radii around the robust radius of each detection, i.e., 0.19 arcsec or 701 pc for detection ${\cal H}$ and 0.04 arcsec or 91 pc for detection ${\cal A}$.

Figure \ref{fig:sidm_comparison} compares our inferred profiles for detections ${\cal H}$ and ${\cal A}$ to projected mass profiles for evolved SIDM haloes with initial NFW halo parameters corresponding to objects at the median inferred redshift of the perturbers and with a concentration set by the median of the redshift–mass–concentration relation given by \cite{Duffy_2008}. We also show projected mass profiles corresponding to the initial NFW model from which each SIDM halo developed. For the more massive detection ${\cal H}$, the mass within the robust radius of 701 pc is reproduced by an SIDM halo with initial NFW values, $\vmax = 77.3$ km/s and $\rmax = 27.3$ kpc (or equivalently, $\mvir = 1.13 \times 10^{11} M_\odot$ and $\cvir = 8.9$). 
However, we note that the virial mass is not a well-defined quantity for subhaloes and that our SIDM treatment does not account for the effect of mass loss due to tidal stripping and is valid only for idealised haloes decoupled from their cosmological context. 

For detection ${\cal A}$, which we have shown to be a field halo at a lower redshift than the lens, we find that an SIDM halo with a CDM analogue of $\mvir = 1.65 \times 10^9 M_\odot$ and $\cvir = 13.3$ has a projected mass within the robust radius of 91 pc that is consistent within $1\sigma$ with the inferred value for this object. However, the slope that is inferred at this radius, $\gamma_{\rm{2D}} = 0.84^{+0.13}_{-0.12}$, is too shallow to be consistent with SIDM at the particular time modeled since our model profile has $\gamma_{\rm{2D}} \geq 1.2$ at all radii. A profile for a slightly higher initial mass halo could fit both the mass and the slope if taken to be at a slightly earlier evolutionary phase when its core radius was still of order 90 pc (rather than much smaller as assumed in Figure \ref{fig:sidm_comparison}).  

For detection $\cal{H}$, our model SIDM halo does reproduce within $1\sigma$ both the enclosed mass within the robust radius and the slope of the enclosed mass profile at this radius. A caveat in both cases is that the gravothermal fluid model is only a good description for isolated haloes in the absence of mergers and baryonic effects. Numerical simulations and semi-analytical models have shown that both tidal stripping and baryonic effects can accelerate the evolution of SIDM haloes, leading to an earlier onset of the core-collapse phase \citep{Nishikawa_2020, Elbert_2018, Sameie_2018, Zhong_2023}. However, the effects of these processes on the mass profiles in the region relevant for our detections is less reliably studied. High-resolution hydrodynamical simulations including a proper treatment of SIDM effects in both the long and the short mean free path regimes are needed to draw more definitive conclusions. This is extremely challenging, particularly since the simulations need to extend through and after the formation of a central black hole. Suitable techniques have not yet been developed. 

Assuming that each halo has evolved over a time interval $\Delta t = t_{\rm{U}} - t_{\rm{L}}(z_{\rm{h}})$, where $t_{\rm{U}}$ is the age of the Universe and $t_{\rm{L}}(z_{\rm{h}})$ is the look-back time at the redshift of each detection, we obtain rough lower limits on the effective cross-section of $\sigma_{c,0}/m_{\rm{dm}} > 200 \ \rm{cm}^2 g^{-1}$ and $\sigma_{c,0}/m_{\rm{dm}} > 300 \ \rm{cm}^2 g^{-1}$ from detection ${\cal H}$ and ${\cal A}$, respectively. If we instead assume a formation redshift of $z_{\rm f}=3$ for the haloes, these limits increase to $\sigma_{c,0}/m_{\rm{dm}} > 300 \ \rm{cm}^2 g^{-1}$ and $\sigma_{c,0}/m_{\rm{dm}} > 400 \ \rm{cm}^2 g^{-1}$, based on ${\cal H}$ and ${\cal A}$, respectively. We stress that these are reduced significantly for haloes that are more concentrated than the median. In the scope of $N$-body simulations with velocity-dependent SIDM cross-sections, \cite{Nadler_2023} have shown that models that reach cross-sections of $\sim 100 \ \rm{cm}^2 g^{-1}$ at the scale of detection ${\cal H}$ can produce core-collapsed (sub)haloes with properties that are qualitatively consistent with those inferred for this object. Moreover, the limits on the effective cross-section are likely to be reduced further both by additional halo compression due to baryonic effects and by the acceleration of evolution due to tidal stripping.

\section{Summary and Discussion}

\label{sec:summary}

We have presented a reanalysis of the gravitational lens systems J0946+1006 and B1938+666 in which the detection of two low-mass haloes, detections ${\cal H}$ and ${\cal A}$ had been previously reported and then later confirmed by independent studies. Our work differs from previous analyses in several ways. Here, we summarise and discuss the implications of our findings.

We find that the data support the presence of $m=1,3,4$ and $m=3,4$ multipoles in the total mass density distribution of the main lens in J0946+1006 and B1938+666, respectively. The strength of the multipoles is different in the two lens galaxies, perhaps indicating different formation histories.

We confirm detection of both low-mass haloes with a statistical significance that is sensitive to the prior on the lens macro-model (i.e. including multipoles or not), to the choice of source regularisation (i.e. gradient or curvature), as well as to the assumed mass density profile of the detected object. Our best-fitting model for J0946+1006 has a detection significance of $\sim$ 10$\sigma$ and $\sim$ 13$\sigma$ for gradient and curvature regularisation, respectively (see Appendix \ref{sec:regularisation}). With curvature source regularisation, a detection significance of $\sim$20$\sigma$ is reached for this system relative to a power-law main lens model without multipoles. For the lens system B1938+666 the best-fitting model is detected with a statistical significance of $\sim$ 11$\sigma$ and $\sim$ 8$\sigma$ for curvature and gradient regularisation.

We confirm the conclusion of previous studies that detection ${\cal H}$ in J0946+1006 is a subhalo of the main lens galaxy with a highly concentrated mass density profile. In contrast to earlier work, however, we find that detection ${\cal A}$ in B1938+666 is a foreground halo at a redshift of $z = 0.12 ^ {+0.07}_{-0.06}$. Its concentration is also high compared to expectations for a CDM halo at this redshift.

For each detection, we have identified a robust radius at which the relative error on the deflection angle and the enclosed mass is minimised; these radii are 0.19 arcsec ($\sim 700$ pc) and 0.04 arcsec ($\sim 90$ pc), for detections ${\cal H}$ and ${\cal A}$ respectively. 
Within a radial range surrounding this radius, both the deflection angle (equivalently the projected mass) and its logarithmic derivative, as defined in Section \ref{sec:modelling_results}, are robustly constrained. All acceptable mass density profiles for the detected object imply values within $1\sigma$ of those found for the best-fit profile.

Comparing the projected mass density profile around the robust radius with the TNG50-1 hydrodynamical simulation, we are able to find analogues of each detected object. In both cases these lie in the high-concentration wings of the distribution of (sub)haloes in the relevant mass and redshift ranges. 
Considering, however, that lensing is inherently more sensitive to high concentration haloes, the preferential detection of such atypical systems should, perhaps, be expected. By matching haloes from the dark-matter-only runs with their counterparts in the TNG50-1, we find that haloes that match the two detections have substantially steeper mass density profiles than their DMO counterperts. Notably, none of the DMO counterparts of the analogues of detection $\cal{H}$ has a density profile that matches the observed one, highlighting the need for baryons to bring the inner density structure of CDM haloes into agreement with the lensing data.

For object $\cal{H}$, the luminosities of matching TNG50 objects are all too high to be consistent with the upper limit on its actual luminosity, even the relatively relaxed upper limit reported by \cite{Minor_2021}. Thus, we
agree with \cite{Minor_2021} and \cite{Despali_2024} that the high concentration of this object is difficult to reconcile with CDM without invoking more baryonic compression than is consistent with its low observed luminosity. This result should nevertheless be interpreted with caution, as only ten host haloes in TNG50-1 have stellar and total masses comparable to those of the J0946+1006 main lens, limiting the statistical power of the result. While simulations with larger physical volumes would provide larger host galaxy samples, they lack the high mass resolution of TNG50 which enables more detailed characterisation of the inner structure of haloes. This is evident in the lower number of subhalo analogues identified by \cite{Minor_2021} in TNG100-1, compared to our analysis and that of \cite{Despali_2024} which are both based on TNG50-1. If detection $\cal{H}$ was accreted at an early time and had its star formation quenched promptly by its host, its low luminosity might perhaps be explained. 

For detection ${\cal A}$, our up-dated, lower redshift implies a new and much lower 3$\sigma$ upper limit on the luminosity,  $L_V < 6.3 \times 10^5 (z/0.13)^2L_{V,\odot}$. In this case, we find that most simulated haloes in TNG50-1 that match the inferred projected mass density profile also satisfy the luminosity limit. Indeed, what differentiates detection ${\cal A}$ from the bulk of the distribution of simulated haloes of similar redshift and total mass is primarily the steep projected mass density slope, which is normally achieved at lower projected masses.

Comparing the projected mass density profiles of the two detections with those predicted near core-collapse for SIDM haloes by the gravothermal fluid model of \cite{Outmezguine_2023}, we find good agreement for detection ${\cal H}$, within $1 \sigma$ in both mass and slope around the robust radius for a halo with initial NFW maximum circular velocity of 77~km/s. For detection ${\cal A}$, the corresponding SIDM halo can reproduce the projected mass at the robust scale but cannot simultaneously match the inferred slope. However, a slightly more massive SIDM halo in a slightly earlier phase of its gravothermal evolution could probably provide an acceptable fit to both the enclosed mass and the projected slope. We have derived a rough lower limit on the velocity-averaged SIDM cross-section of $\sigma_{c,0}/m_{\rm{dm}} > 300 \ \rm{cm}^2 g^{-1}$. Note that the characteristic velocities in the two cases differ by a factor of around three, so it may be difficult to ensure that the core is simultaneously in an advanced stage of collapse in both cases.

Our results suggest that velocity-dependent SIDM models could, in principle, provide a plausible explanation for the observed properties of the detected haloes. Interestingly, for detection $\cal{H}$ both the mass within a 300 pc radius and the luminosity are comparable to those of M32, which is not only the most concentrated known dwarf galaxy in the local Universe (and hence the most likely nearby site for core collapse to have occurred) but also hosts a central black hole with a well-measured mass of $M_\bullet = 2.5 \times 10^6 M_\odot$ \citep{Verolme_2002}. This is surprisingly large for a galaxy of luminosity just $\sim 1.5\times 10^8L_\odot$ and corresponds approximately to the black hole mass expected in a core-collapsed SIDM halo, if the seed black hole promptly accretes the inner core which is in the short mean free path (SMFP) regime, and thereafter undergoes minimal growth. 

We note that the adopted model in Section \ref{sec:compare_with_sidm} only applies to idealised spherically symmetric haloes decoupled from any cosmological context and in the absence of baryons. All these effects have been shown to play an important role in the evolution of SIDM haloes and specifically on the fraction of core-collapsed objects of different masses. High-resolution hydrodynamical simulations incorporating a broad range of velocity-dependent models and able to integrate both the long and short mean free path regimes into the post-black hole formation epoch are necessary for a proper quantitative comparison between SIDM predictions and the inferred properties of detections ${\cal H}$ and ${\cal A}$. Taken together with the observational evidence from ultra-diffuse galaxies (UDGs), which require significantly lower concentrations than those predicted by CDM to reproduce their observed rotation curves (e.g., \citealp{Guo_2019, Mancera_2019, Mancera_2021}; though see \citealp{Banik_2022b, Sellwood_2022}) and observed level of tidal susceptibility for the dwarf galaxies in the Fornax cluster \citep{Asencio_2022}, such simulations would provide an interesting avenue to explore whether SIDM models can accommodate the full range of central density profiles observed, which remains difficult to reconcile within the CDM framework.

\section*{Acknowledgments}
S.~V. thanks the Max Planck Society (MPG) for support through a Max Planck Lise Meitner Group. This research was carried out on the High-Performance Computing resources of the Freya and Raven clusters at the Max Planck Computing and Data Facility (MPCDF) in Garching, operated by the MPG. J.~P.~M. acknowledges support from the National Research Foundation of South Africa (Grant Number: 128943). G.~D. acknowledges the funding by the European Union - NextGenerationEU, in the framework of the HPC project – “National Centre for HPC, Big Data and Quantum Computing” (PNRR - M4C2 - I1.4 - CN00000013 – CUP J33C22001170001).

\section*{Data Availability}

The observational data used in this paper are publicly available through the HST and Keck archives. The TNG50 simulation data are accessible at \url{https://www.tng-project.org} \citep{Nelson_2019a}.

\bibliographystyle{mnras}
\bibliography{ms}


\appendix

\section{\texorpdfstring{Comparison with Sengül et al.\ 2022}{Comparison with Sengul et al. 2022}}

\label{app:NIC1}

\begin{figure*}
\includegraphics[width=\textwidth]{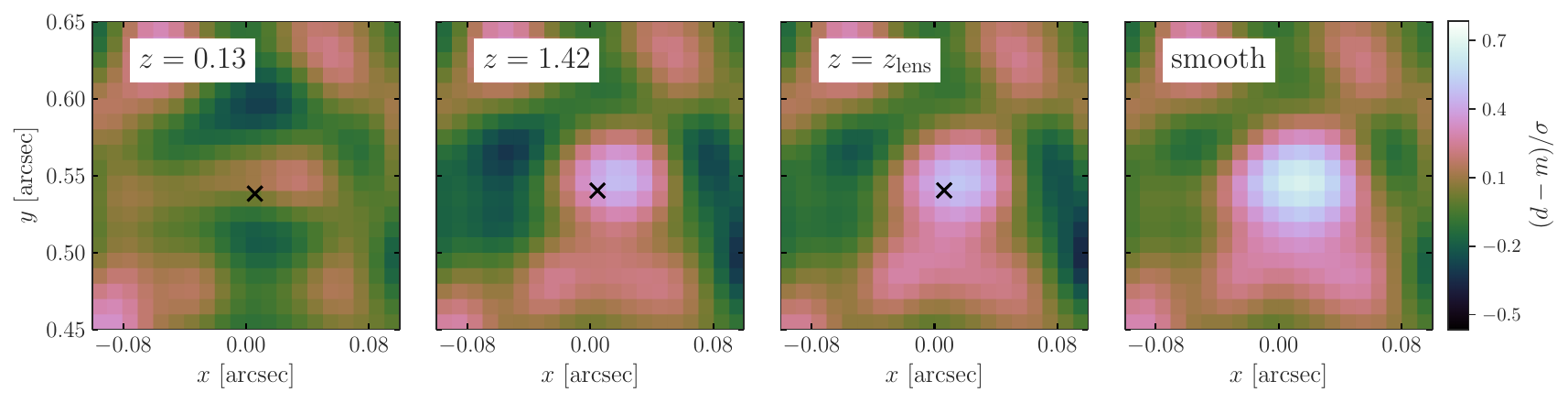}
    \caption{Normalized image residuals for three different redshift values of detection $\cal{A}$ (first three panels) and that of a smooth model (right panel) that does not include any low-mass halo. All residuals are smoothed with a Gaussian kernel to highlight differences in the performance of each model.}
\label{fig:residuals_z}
\end{figure*}

\begin{figure}
	\includegraphics[width=0.5\textwidth]{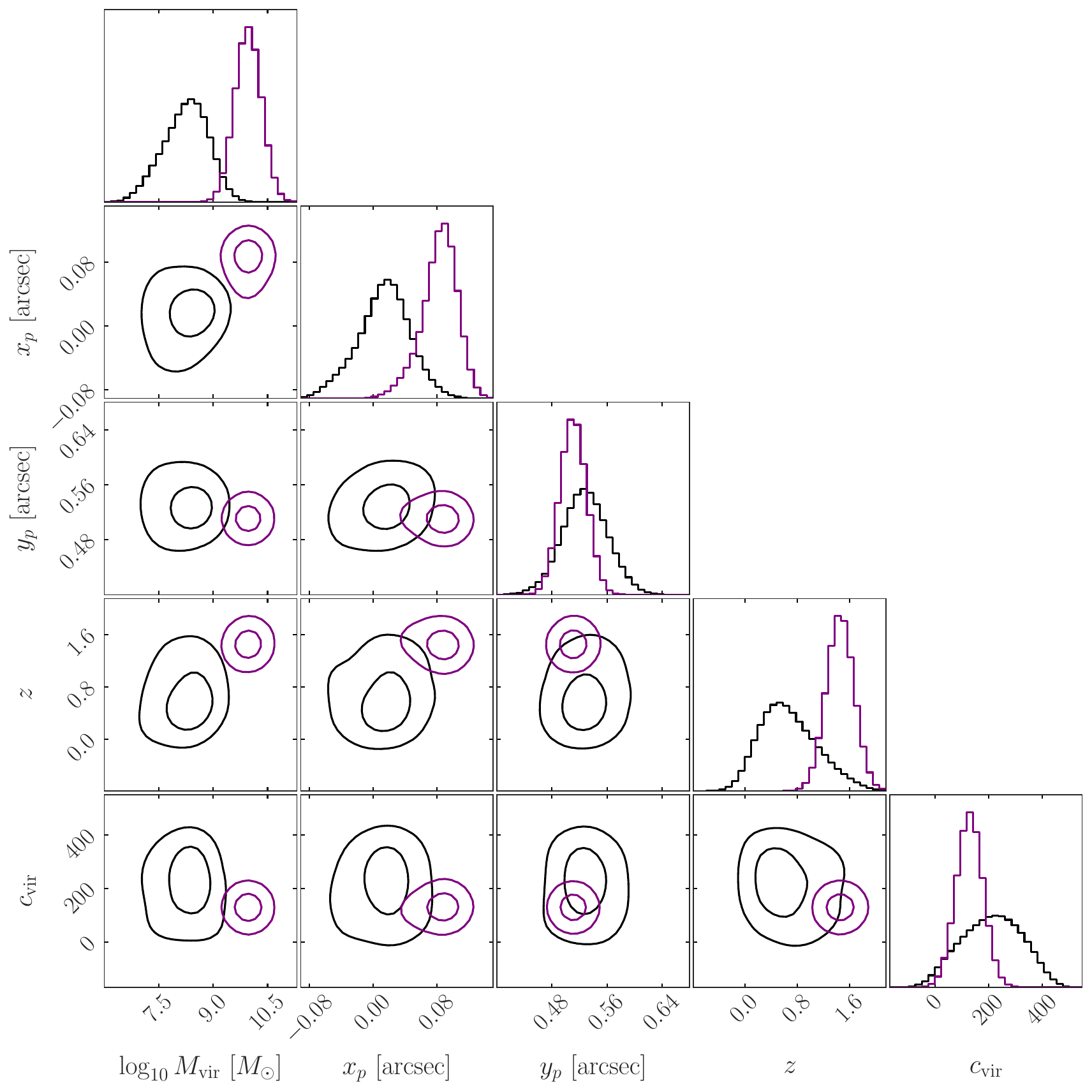}
    \caption{Posterior distributions for the parameters of detection ${\cal A}$ obtained from modelling the HST NIC1 data assuming the correct PSF model (black curves) and an unphysical, $\delta$-function model (purple curves).}
    \label{fig:NIC1}
\end{figure}

As a reminder, our analysis is based on the imaging obtained with the Keck-II telescope using an AO system, rather than on the HST NICMOS/NIC1 data used by \citet{Seng_l_2022}. This is because the Keck AO data have better angular resolution and higher SNR than the NIC1 data, both of which lead to better sensitivity to low-mass haloes \citep{Despali_2022}. The NIC1 camera has a native pixel scale of 0.043 arcsec and an angular resolution of $\sim$0.15 arcsec for the F160W filter. Although the native pixel scale properly samples the PSF, the data provided by the HST Legacy Archive has been drizzled to a pixel scale of 0.025 arcsec. For the Keck AO data, the angular resolution is variable, but the SHARP data typically achieve resolutions of 0.06--0.09 arcsec for the $K^\prime$ filter. The pixel scale of the camera is 0.01 arcsec.  The improvement in resolution provided by the AO observations is clearly detectable by eye when comparing to the HST NICMOS data. The HST and Keck images have similar SNR per pixel values in the region of the image where object ${\cal A}$ is detected, but given the smaller AO pixels this is not a meaningful comparison of sensitivity. 
In order to compare sensitivities more appropriately, we bin both the HST/NICMOS and Keck AO data to a pixel scale of 0.05 arcsec, finding that the AO data have an SNR that is a factor of 3 higher than the HST data.

In Fig. \ref{fig:residuals_z}, we show the normalised image residuals in a region around the location of the detected object assuming three different values of its redshift, when modelling the Keck data. This plot, combined with the values of Bayesian evidence in Table \ref{tab:B1938_pert} clearly indicates that the data analysed in this paper provide enough information to constrain the redshift of detection $\cal{A}$. On the other hand, we find that the HST NIC1 data do not allow us to reach any definitive conclusion either on the redshift of the considered object or, indeed, on its presence. 

To enable a direct comparison with the results of \citet{Seng_l_2022}, we model the same data assuming a power-law model plus external shear for the mass distribution of the main lens galaxy. We find that the Bayesian evidence for a model including detection $\cal{A}$ is only marginally higher than that of a smooth model, with $\Delta\log{\cal{E}} = 1.3$, showing no statistically significant preference for the perturbed model in this dataset. Notably, the Keck AO data yield a $\Delta\log{\cal{E}}$ of at least 42.4.

The inferred virial mass and concentration of the detected object are consistent within $1\sigma$ between the two datasets. However, the redshift inferred from the HST data is $z = 0.61^{+0.48}_{-0.34}$, remaining effectively unconstrained over almost the full range between the observer and the source (see posterior distributions in Fig. \ref{fig:NIC1}). We conclude, therefore, that the HST NIC1 data lack the constraining power to reliably determine the redshift of detection $\cal{A}$. This is in contrast to the results of \citet{Seng_l_2022}, who inferred a redshift of $z = 1.42^{+0.10}_{-0.15}$ for their NFW field halo model, with a significant increase in the Bayesian evidence of $\Delta\log_{10}{\cal{E}} = 18.0$ relative to their smooth model. 

We find that if the NIC1 data are modelled with an unphysical delta-function PSF, the perturbed model yields a substantial increase in log-evidence of $\Delta\log{\cal{E}} = 25.4$ relative to the smooth model. In this case, all perturber parameters become more tightly constrained, with detection $\cal{A}$ placed at a redshift of $z = 1.45^{+0.10}_{-0.10}$, consistent with the results of \citet{Seng_l_2022}. This may, perhaps, shed some light on the difference between our analysis and the results they quote.

\section{Effect of source regularisation}
\label{sec:regularisation}

\cite{Ballard_2024} have found the detection significance of $\cal{H}$ to be highly sensitive to the source regularisation. While this may seem problematic, it is to be expected. Lensing is an ill-posed, under-constrained problem and priors on any of the unknown quantities, i.e. lens mass distribution, light profile or source model, are bound to play an important role. It is well established that, within the limits of the SNR of the data and their angular resolution, complexity in the source can be partially degenerate with complexity in the lensing potential  \citep[e.g.][]{Vernardos_2022}.

For the specific case of object $\cal{H}$, \cite{Minor_2025} have shown that a "checkerboard" pattern appears in the reconstructed source when gradient regularisation is used without additional complexity in the main lens, such as multipoles or substructure. This effect artificially reduces the image-plane residuals, potentially masking the need for additional complexities in the macro-model; the residuals remain similar even when a subhalo or multipoles are added, offering at best a small improvement in Bayesian evidence, mostly related to the source being smoother. However, when supersampling is used, the reconstructed source becomes smoother even for a smooth model with gradient regularisation, and the significance of subhalo detection becomes robust to different choices of source prior.

\begin{figure*}
    \centering
\includegraphics[width=1.0\textwidth]{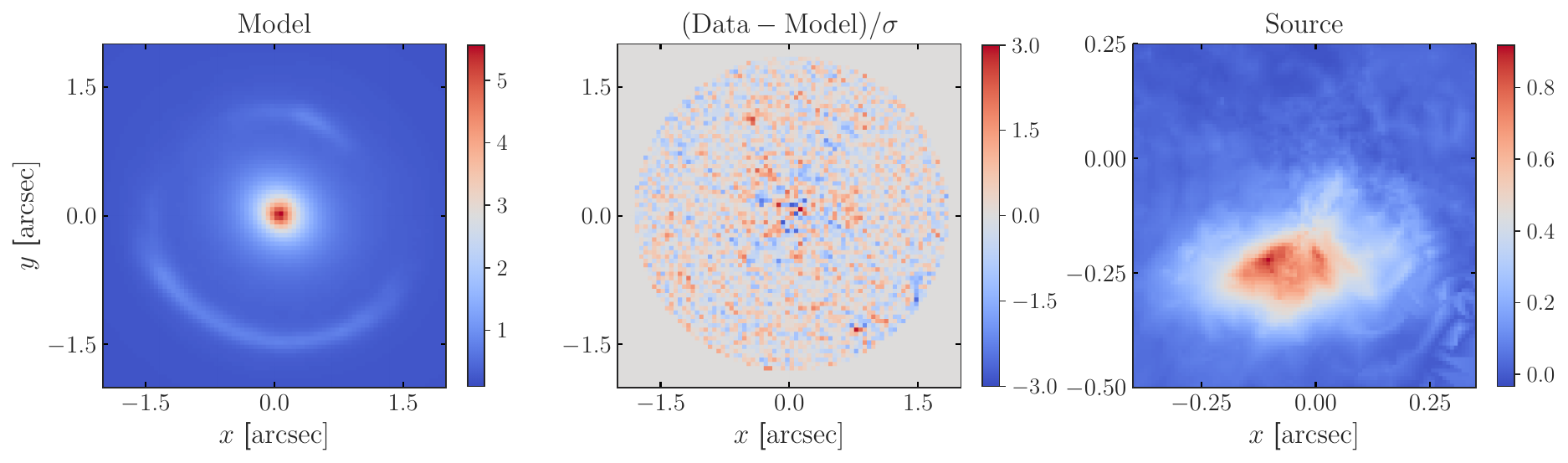}
    \caption{From left to right: the model, normalised image residuals and reconstructed source with the gradient regularisation, for the MAP parameters of the smooth model excluding the subhalo and multipoles for the gravitational lens system J0946+1006.}
    \label{fig:DR_RES_GRAD_SMOOTH}
\end{figure*}

In contrast to the results presented in \cite{Ballard_2024} and \cite{Minor_2025}, we do not recover noise-level residuals for the smooth model when using gradient regularisation, even without supersampling. Instead, a clear, localised residual at the position of the perturber is evident in Fig. \ref{fig:DR_RES_GRAD_SMOOTH}, highlighting the need for additional mass complexity in the model. 
The increase in Bayesian evidence when multipoles and/or a perturber are included in models with gradient-regularised sources is not merely a result of smoother source reconstruction, but also reflects a clear improvement in fit quality and image-plane residuals. Indeed, without supersampling, we find an increase in the Bayesian evidence of $\Delta\log{\cal{E}} = 75.6$ for the gradient-regularised model that includes an NFW subhalo (without multipoles) over a smooth model that does not. 

These values are significantly higher than reported by \cite{Ballard_2024} and \cite{Minor_2025}, who find $\Delta\log{\cal{E}} = 7.2$ and $\Delta\log{\cal{E}} = 16.3$, respectively. The lens light in this system has relatively intricate structure, and the light residuals presented in \cite{Ballard_2024} and \cite{Minor_2025} appear noticeably stronger than our own. 
This suggests that the checkerboard effect and the inconsistencies in reported detection significances are not due solely to differences in the implementation of source regularisation, but also to differences in the modelling, data masking and foreground subtraction.
Even though we do not observe the issues reported by \cite{Minor_2025} in our own analysis, we still find that curvature regularisation leads to a more significant detection of the perturber. 
In this case, including the NFW subhalo results in a log-evidence increase of 81.6 for the PL$_{\rm lens}$+$\rm MP_{1,3,4}$+$\cal H$ model relative to the PL$_{\rm lens}$+$\rm MP_{1,3,4}$, and an increase of 195.9, when comparing the PL$_{\rm lens}$+$\cal H$ to the baseline PL model.
We note that our highest detection significance is even stronger than reported by \cite{Minor_2025}, however, we caution that, as we model different data (we only consider the lowest-redshift source) a direct comparison of evidence values and Bayes factors may not be valid.

\begin{figure*}
    \centering
\includegraphics[width=1.0\textwidth]{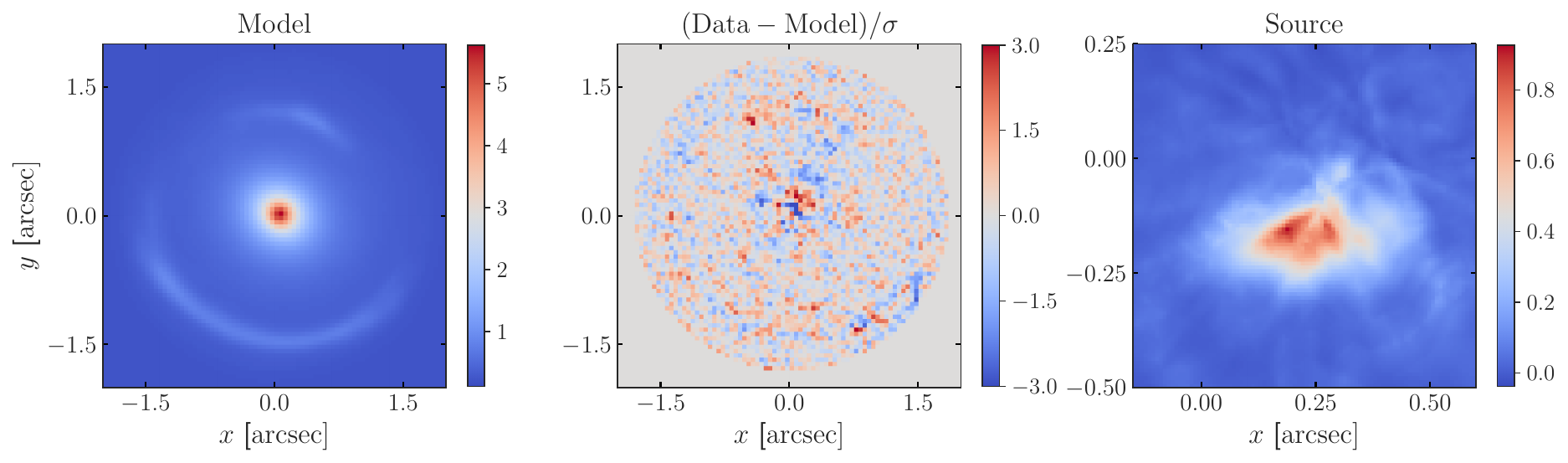}
    \caption{Same as \ref{fig:DR_RES_GRAD_SMOOTH}, but with supersampling performed on the image pixels.}
    \label{fig:SUPER_DR_SMOOTH}
\end{figure*}

To test the impact of supersampling, we repeated the analysis for both the smooth and best‑fitting PL$_{\rm lens}$+$\rm MP_{1,3,4}$+$\cal H$ models with gradient-regularised sources, splitting each image pixel into $4\times4$ subpixels and modelling the lens light simultaneously with the mass distribution. As shown in Figure \ref{fig:SUPER_DR_SMOOTH}, the checkerboard effect is no longer present in the reconstructed source when supersampling is performed. Upon including the NFW subhalo and multipoles, we find an increase in the Bayesian evidence of $\Delta\log{\cal{E}} = 86.0$ relative to the baseline PL$_{\rm lens}$ model. Although supersampling leads to a modest enhancement in the detection significance of the perturber, the overall boost remains smaller than that reported by \cite{Minor_2025}, likely due to differences in lens light treatment. Figure \ref{fig:DR_mn_sub_vmax} shows the posterior distributions of the subhalo parameters from the supersampled run. The inferred parameters remain consistent within $1\sigma$ of those obtained in the main analysis without supersampling. The projected mass and slope at the robust radius are $M_{\rm{2D}}(\theta_{\cal{H}}) = 2.76^{+0.15}_{-0.14}\times10^{9} M_\odot$ and $\gamma_{\rm{2D}}(\theta_{\cal{H}})=1.11^{+0.06}_{-0.07}$, respectively, leaving our main conclusions for object $\mathcal{H}$ effectively unchanged.

Adopting a gradient regularisation scheme for the background source galaxy in the lens system B1938+666, we find a reduction in Bayesian evidence. $\Delta\log{\cal{E}} = -25.2$ compared to the curvature-regularised smooth model. This is before including any macro-lens multipoles or low-mass perturber and suggests that the former scheme is less suitable for modelling the source surface brightness in this dataset. 
Comparing the PL$_{\rm lens}$+$\rm MP_{3,4}$+$\cal A$ model, where object $\cal{A}$ is modelled as a PL field halo and the source is regularised with a gradient prior, to the PL$_{\rm lens}$+$\rm MP_{3,4}$ model, yields a log-evidence difference of $\Delta\log{\cal{E}} = 31.6$. When multipoles are not included, the PL$_{\rm lens}$+$\cal A$ model yields a log-evidence difference of $\Delta\log{\cal{E}}_{3,4} = 34.1$ relative to the baseline PL$_{\rm lens}$.
Thus, even when including detection $\mathcal{A}$ and multipoles, the Bayesian evidence is still lower for gradient regularisation than for curvature regularisation. 

For both systems, we find that, regardless of changes in the statistical significance of the detection, the properties of the main deflector are robustly constrained and remain consistent within $\sim1\sigma$ between the two regularisation schemes (see Figs. \ref{fig:macro_J0946} and \ref{fig:macro_b1938}). Moreover, the inferred redshift and mass profile parameters of the best-fitting models for the two detections are also consistent within $1\sigma$, as shown in Figs. \ref{fig:DR_mn_sub_vmax} and \ref{fig:b1938_mn_pl}.

\section{Lens modelling results}
\label{app:posteriors}

\begin{figure*}
    \centering
    \includegraphics[width=1.0\textwidth]{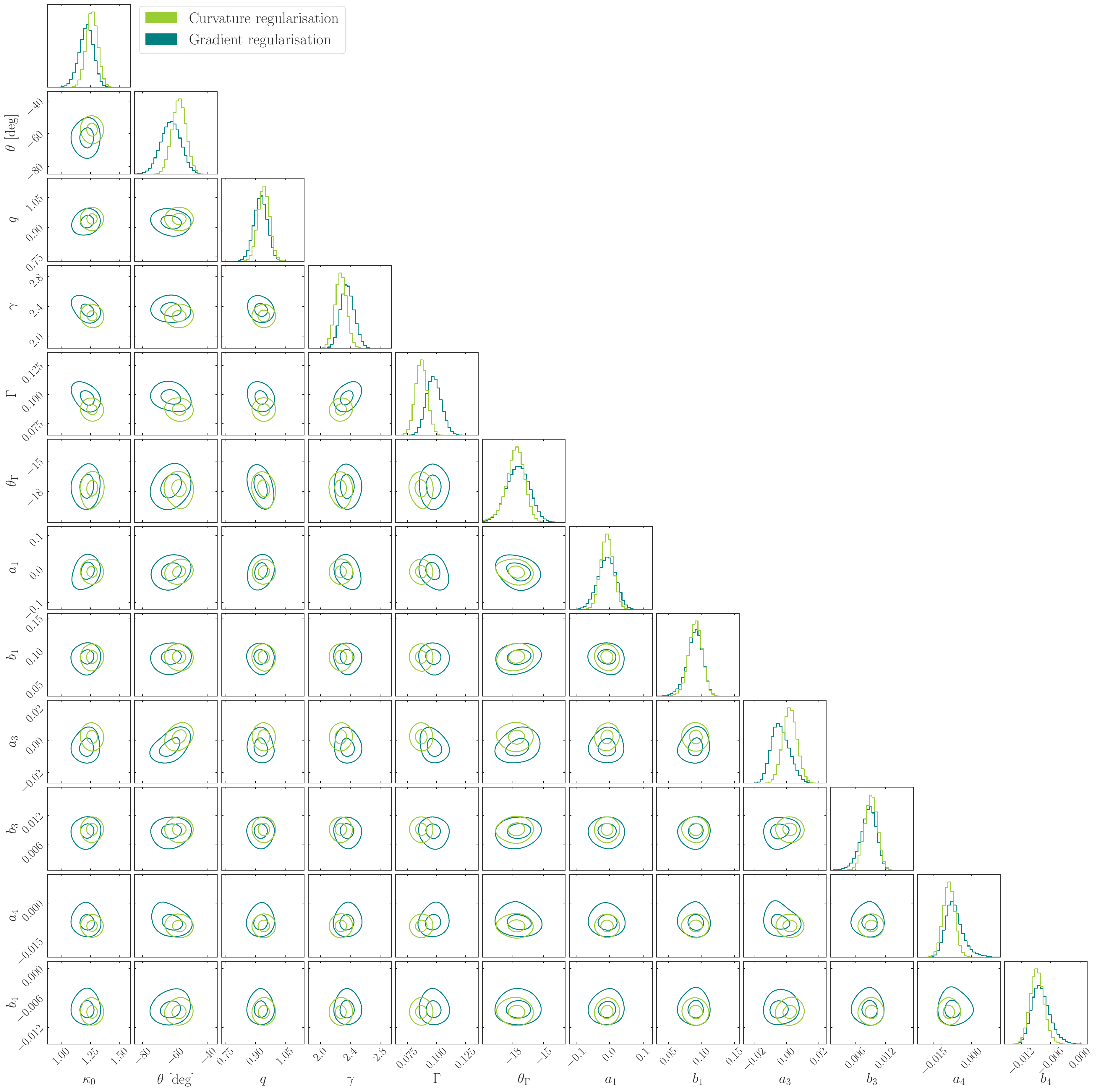}
    \caption{Posterior distributions of the macro-model parameters for the gravitational lens system J0946+1006 for two types of source regularisation.}
    \label{fig:macro_J0946}
\end{figure*}

\begin{figure*}
    \centering
    \includegraphics[width=1.0\textwidth]{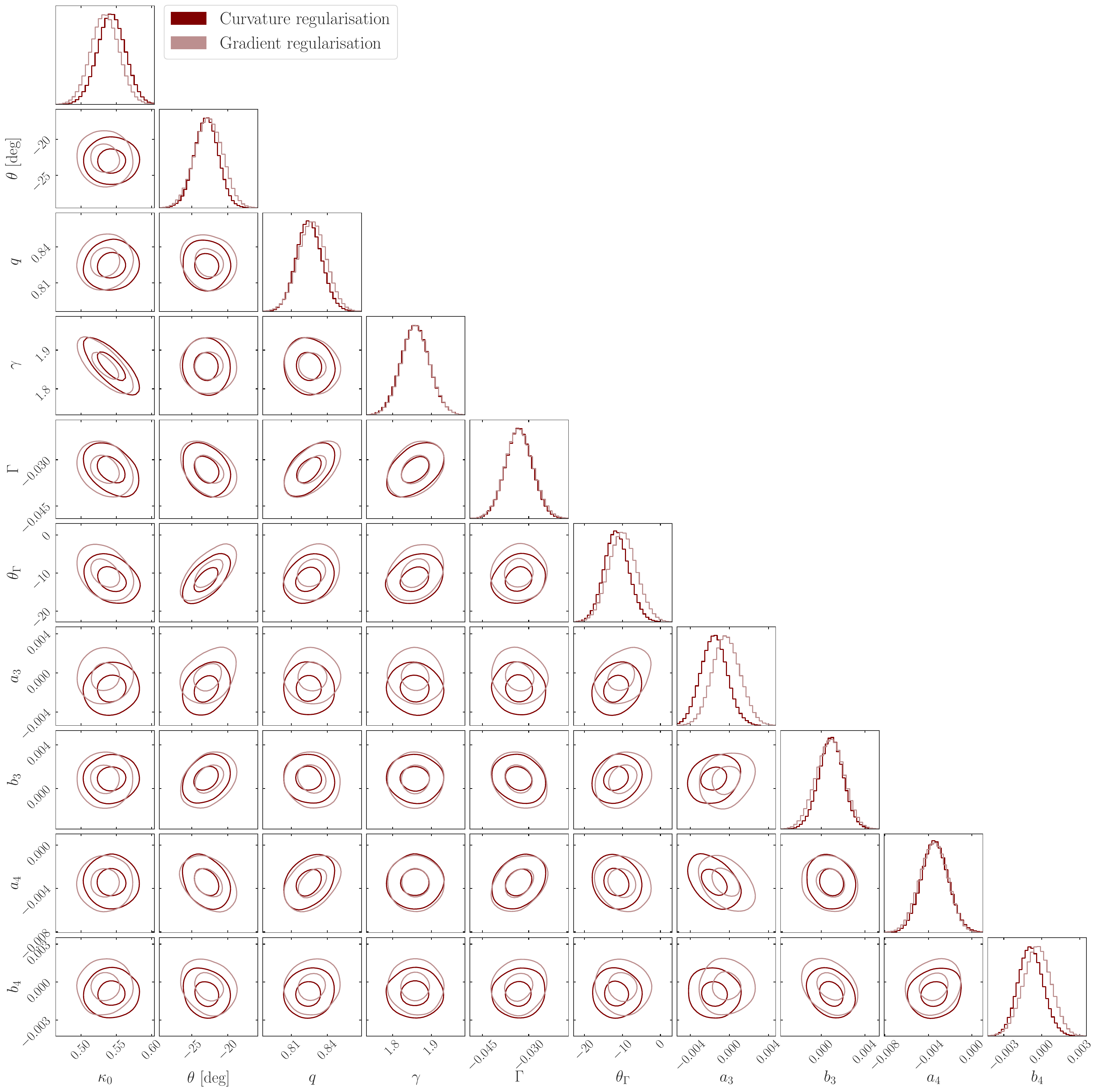}
    \caption{Same as \ref{fig:macro_J0946}, but for the gravitational lens system B1938+666.}
    \label{fig:macro_b1938}
\end{figure*}

\begin{figure*}
    \centering
     \begin{minipage}{0.49\textwidth}
        \centering
        \includegraphics[width=\textwidth]{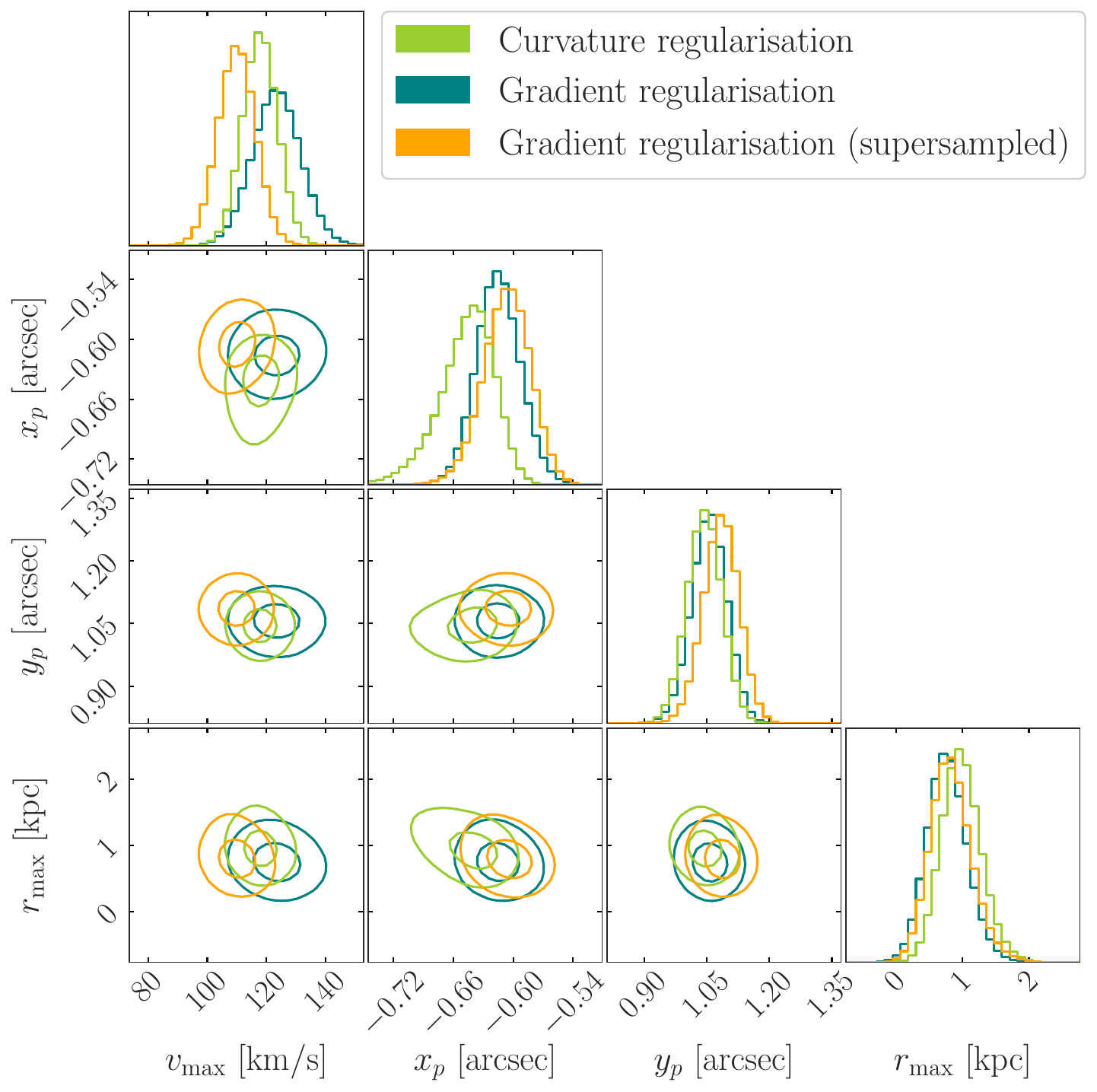}
        \caption{Posterior distributions for the parameters of detection ${\cal H}$ when modelled as an NFW subhalo, i.e. $z\equiv z_{\rm lens}$.}
        \label{fig:DR_mn_sub_vmax}
    \end{minipage}
    \hfill
    \begin{minipage}{0.49\textwidth}
        \centering
        \includegraphics[width=\textwidth]{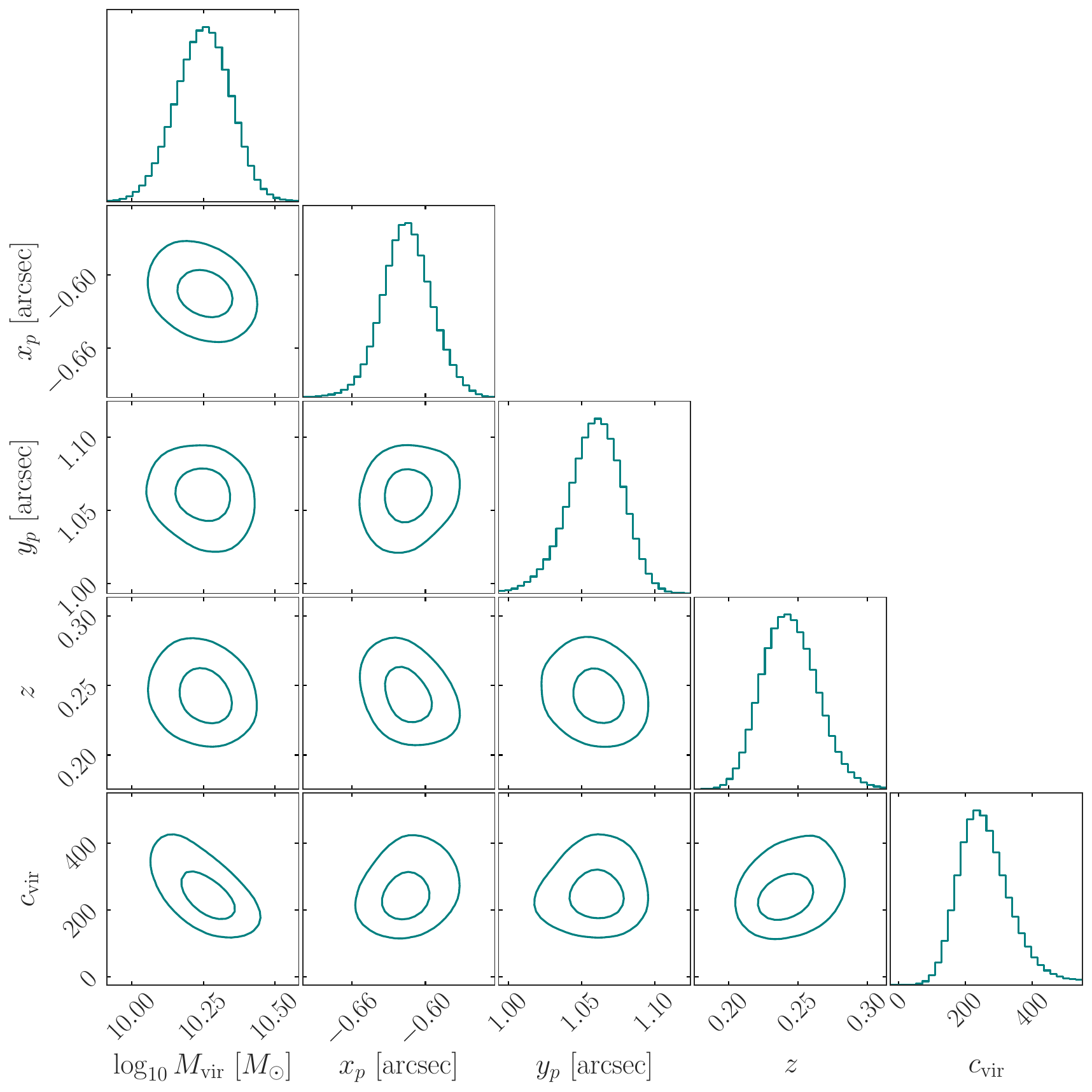}
        \caption{Same as Fig. \ref{fig:DR_mn_sub_vmax} for an NFW profile with free redshift and concentration.}
        \label{fig:DR_mn_NFW}
    \end{minipage}
\end{figure*}

\begin{figure*}
    \centering
     \begin{minipage}{0.49\textwidth}
        \centering
         \includegraphics[width=\textwidth]{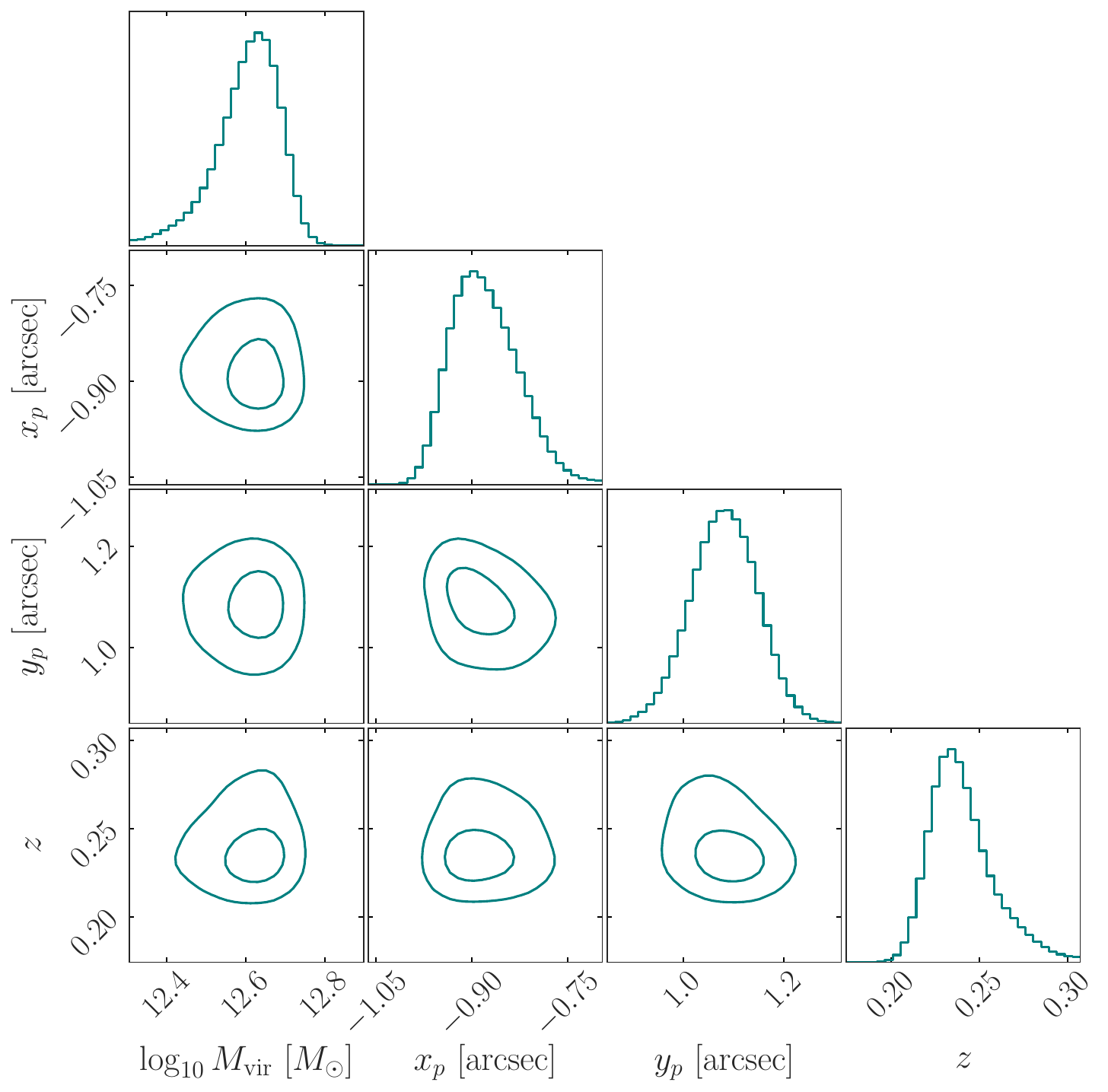}
        \caption{Same as Fig. \ref{fig:DR_mn_sub_vmax} for an NFW profile with free redshift and concentration taken from the median of the \citet{Duffy_2008} mass-concentration-redshift relation.}
        \label{fig:DR_mn_NFW_fix}
    \end{minipage}
    \hfill
    \begin{minipage}{0.49\textwidth}
         \centering
         \includegraphics[width=\textwidth]{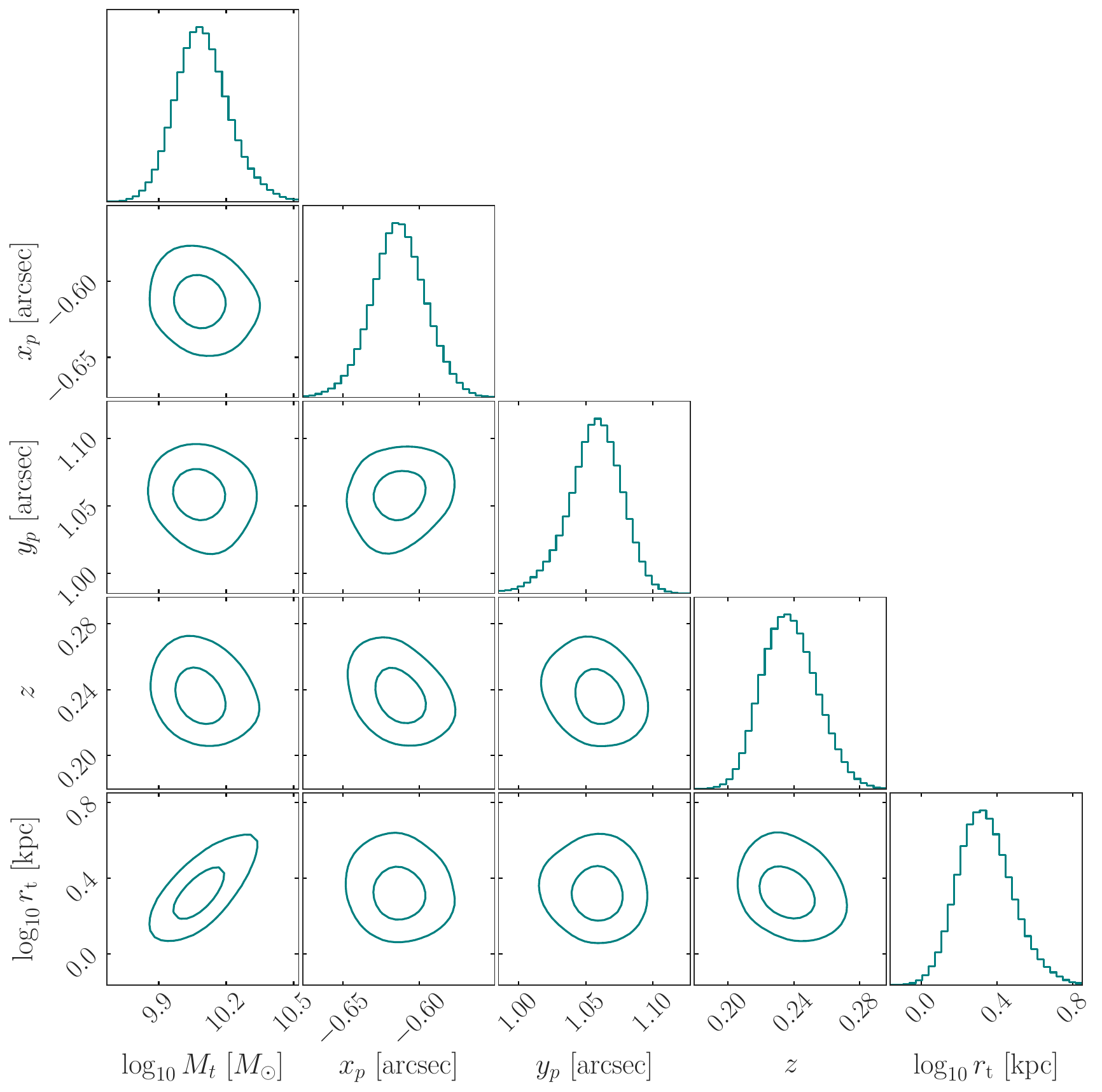}
         \caption{Same as Fig. \ref{fig:DR_mn_sub_vmax} for the pseudo-Jaffe profile.}
        \label{fig:DR_mn_PJ}
    \end{minipage}
\end{figure*}

\begin{figure*}
    \centering
     \begin{minipage}{0.49\textwidth}
        \centering
        \includegraphics[width=\textwidth]{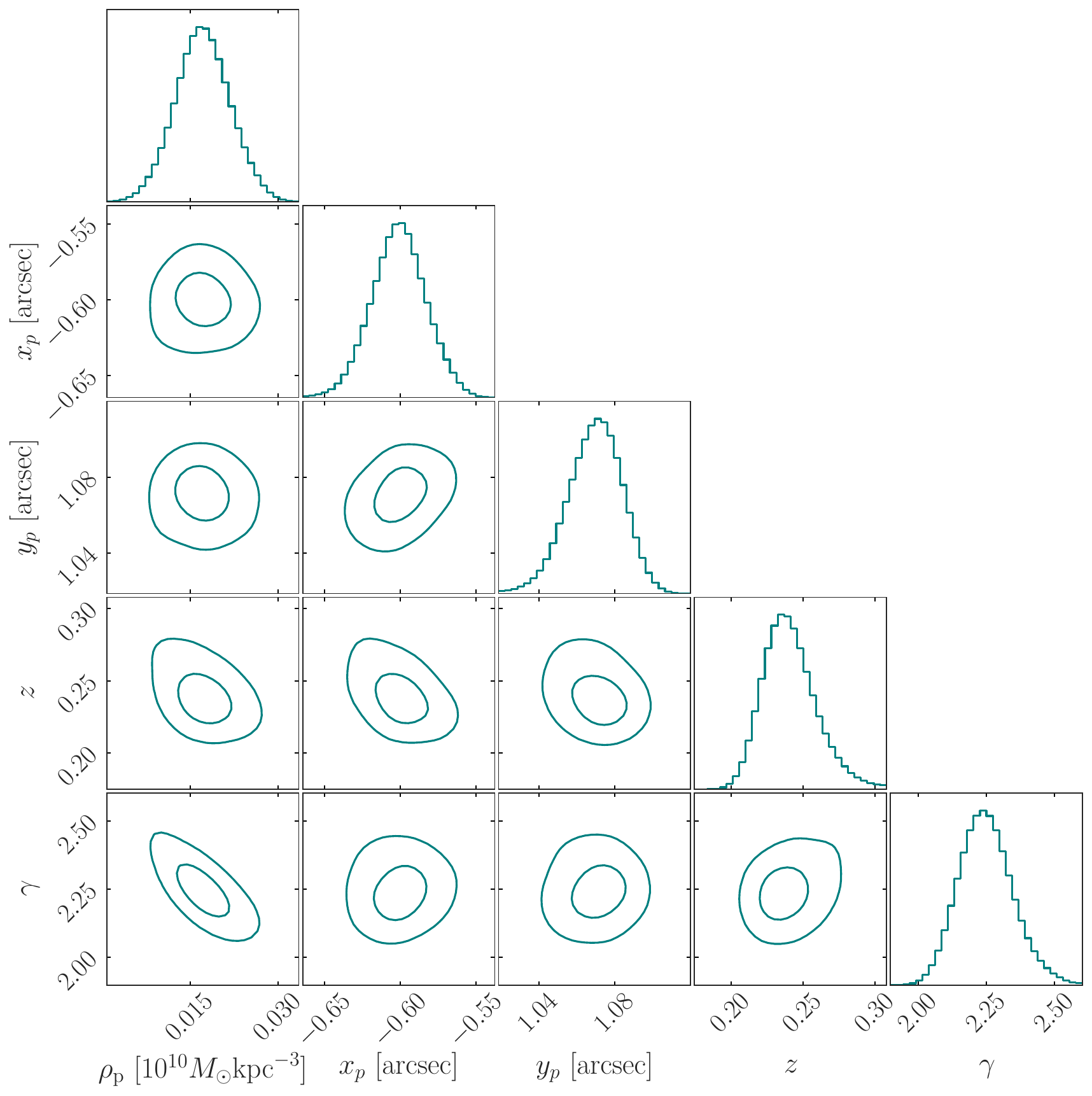}
       \caption{Same as Fig. \ref{fig:DR_mn_sub_vmax} for the PL profile.}
         \label{fig:DR_mn_pl}
    \end{minipage}
    \hfill
    \begin{minipage}{0.49\textwidth}
        \centering
        \includegraphics[width=\textwidth]{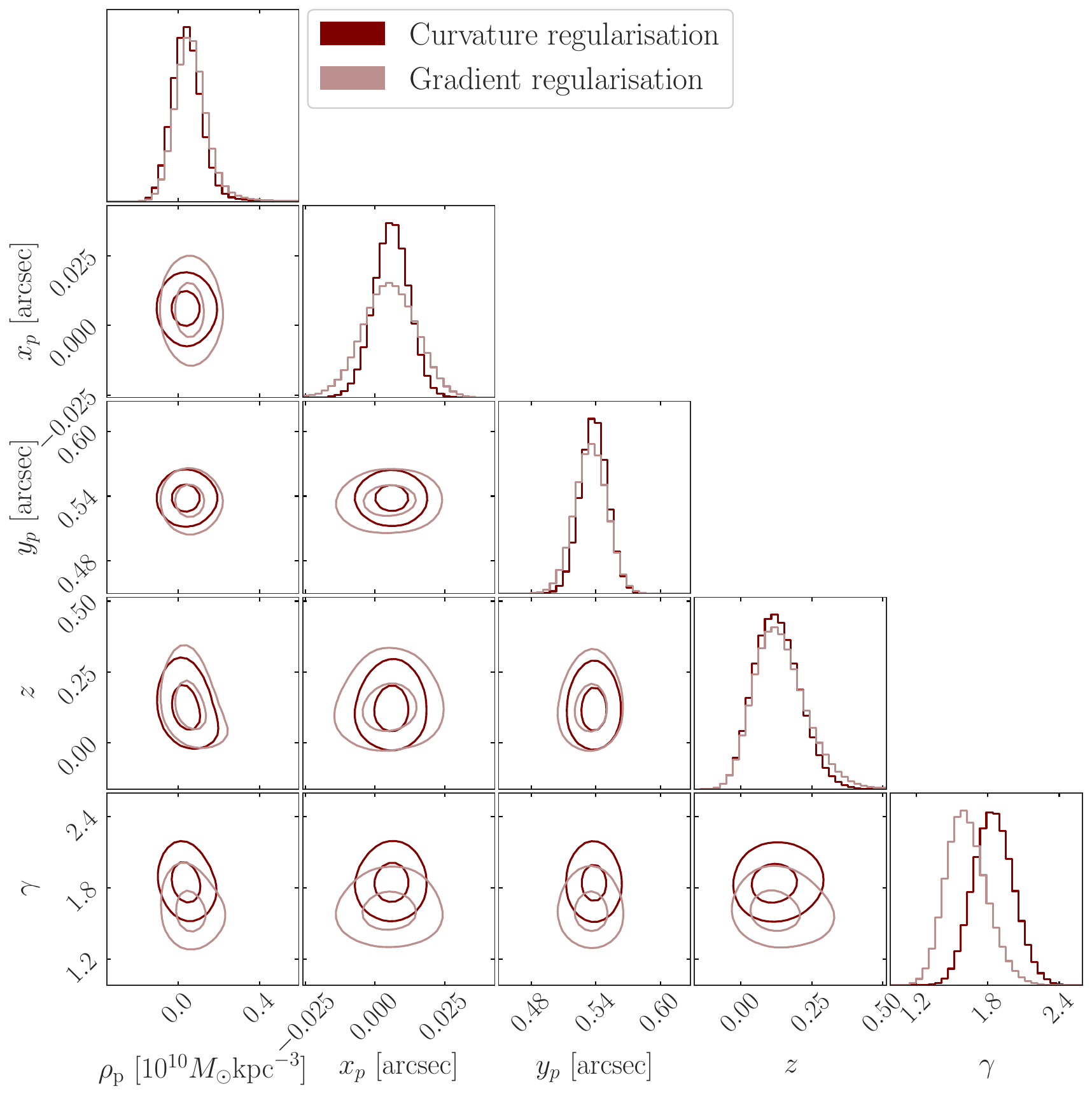}
        \caption{Posterior distributions for the parameters of detection ${\cal A}$ when modelled with a PL profile.}
        \label{fig:b1938_mn_pl}
    \end{minipage}
\end{figure*}

\begin{figure*}
    \centering
     \begin{minipage}{0.49\textwidth}
        \centering
        \includegraphics[width=\textwidth]{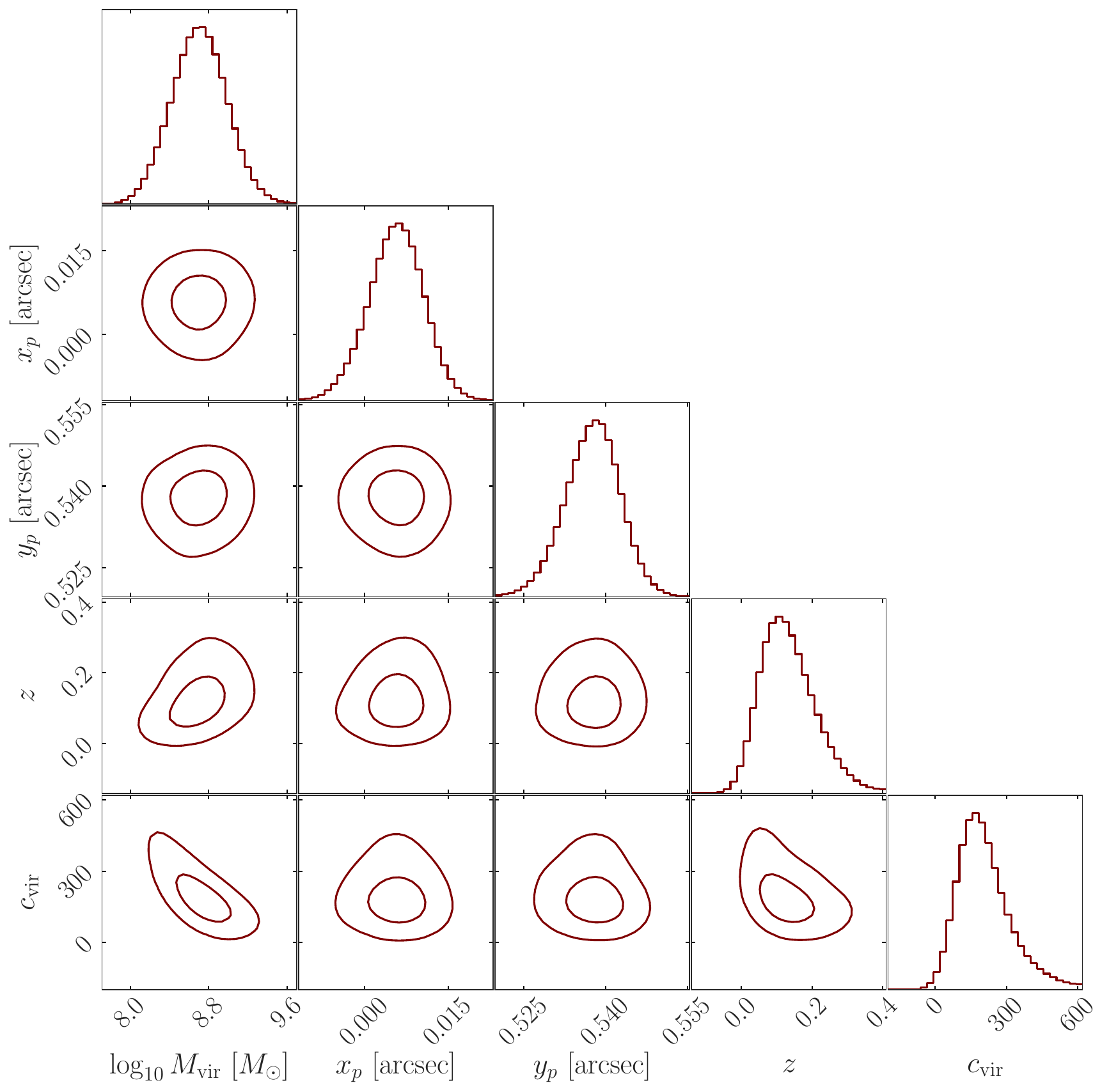}
        \caption{Same as Fig. \ref{fig:b1938_mn_pl} for an NFW halo.}
        \label{fig:b1938_mn_mvir}
    \end{minipage}
    \hfill
    \begin{minipage}{0.49\textwidth}
        \centering
       \includegraphics[width=\textwidth]{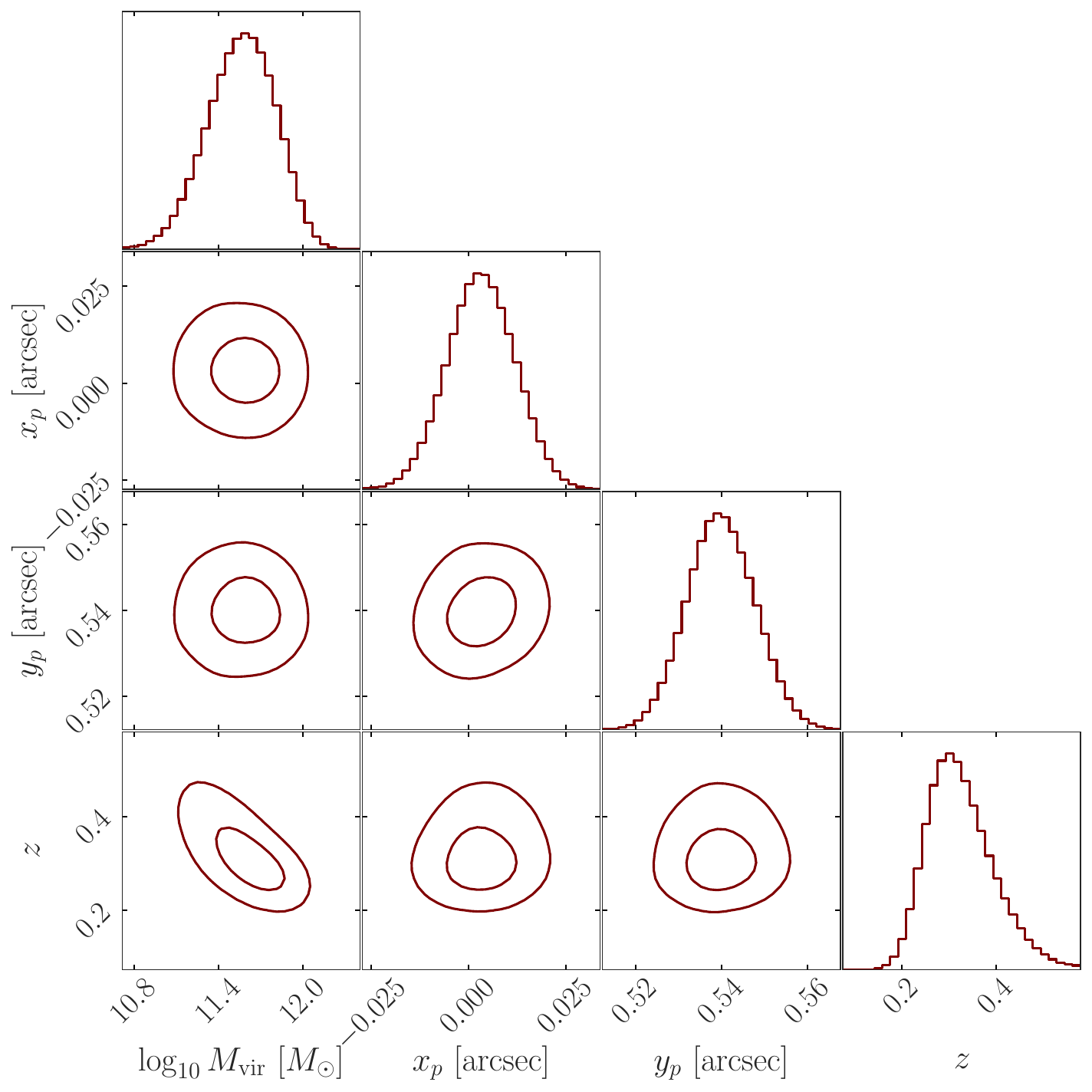}
    \caption{Same as Fig. \ref{fig:b1938_mn_pl} for an NFW profile with concentration taken from the median of the \citet{Duffy_2008} mass-concentration-redshift relation.}
    \label{fig:b1938_mn_duffy}
    \end{minipage}
\end{figure*}

\begin{figure*}
    \centering
\includegraphics[width=1.0\textwidth]{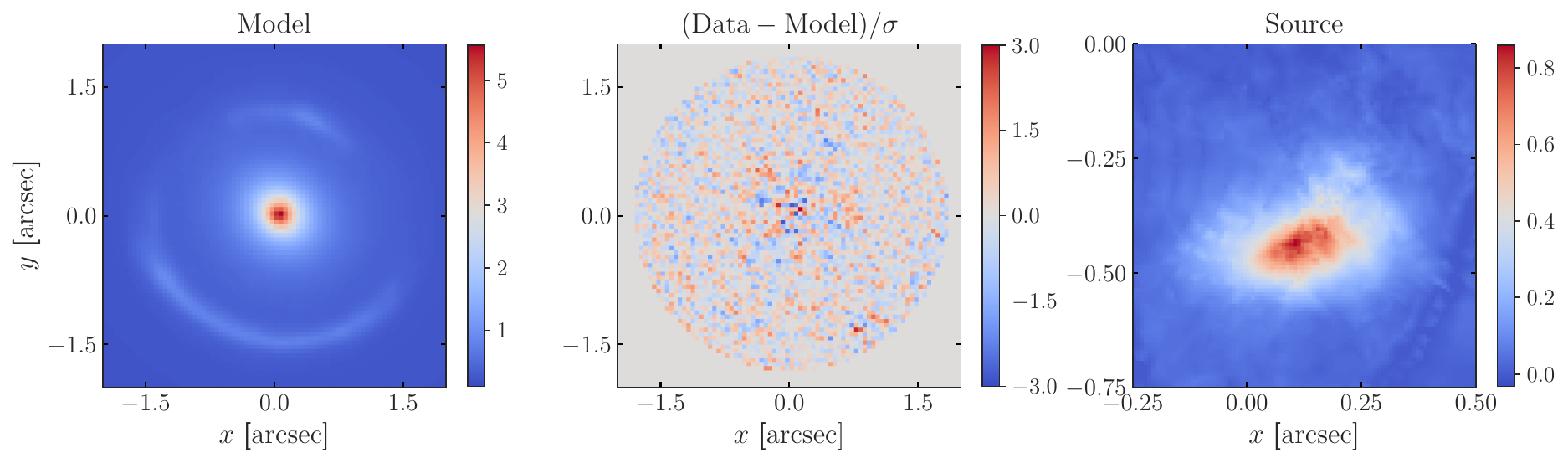}
    \caption{Same as \ref{fig:DR_RES_GRAD_SMOOTH}, but for the MAP parameters of the best-fitting model that includes the NFW subhalo and multipoles.}
    \label{fig:DR_RES}
\end{figure*}

\begin{figure*}
    \centering
\includegraphics[width=1.0\textwidth]{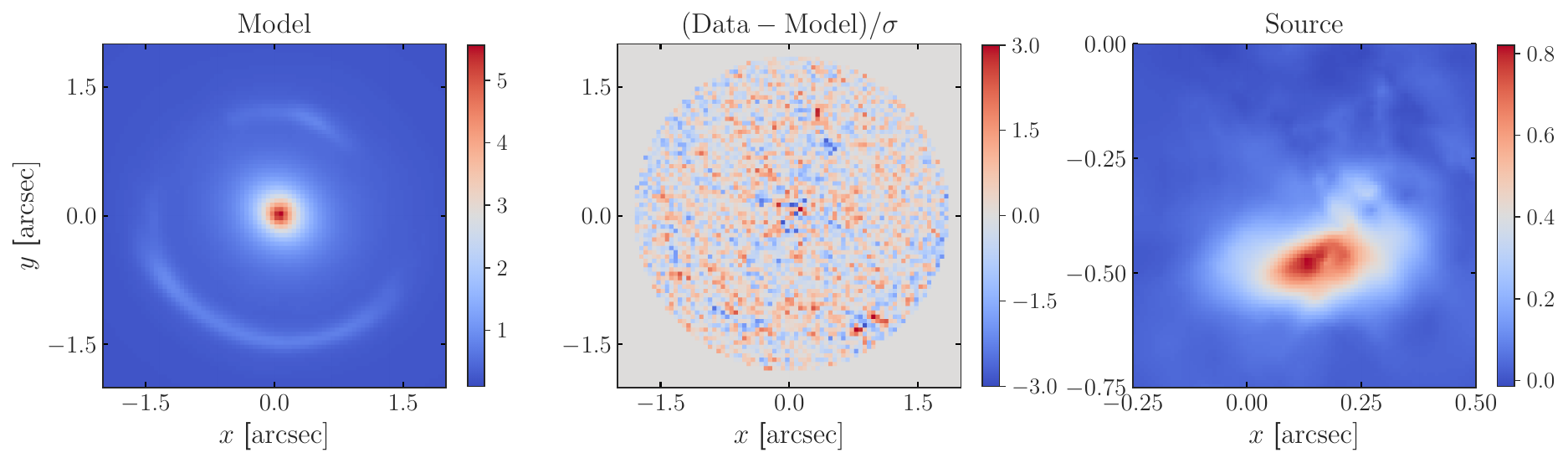}
    \caption{Same as \ref{fig:DR_RES}, but with the curvature regularisation.}
    \label{fig:DR_RES_CURV}
\end{figure*}

\begin{figure*}
    \centering
\includegraphics[width=1.0\textwidth]{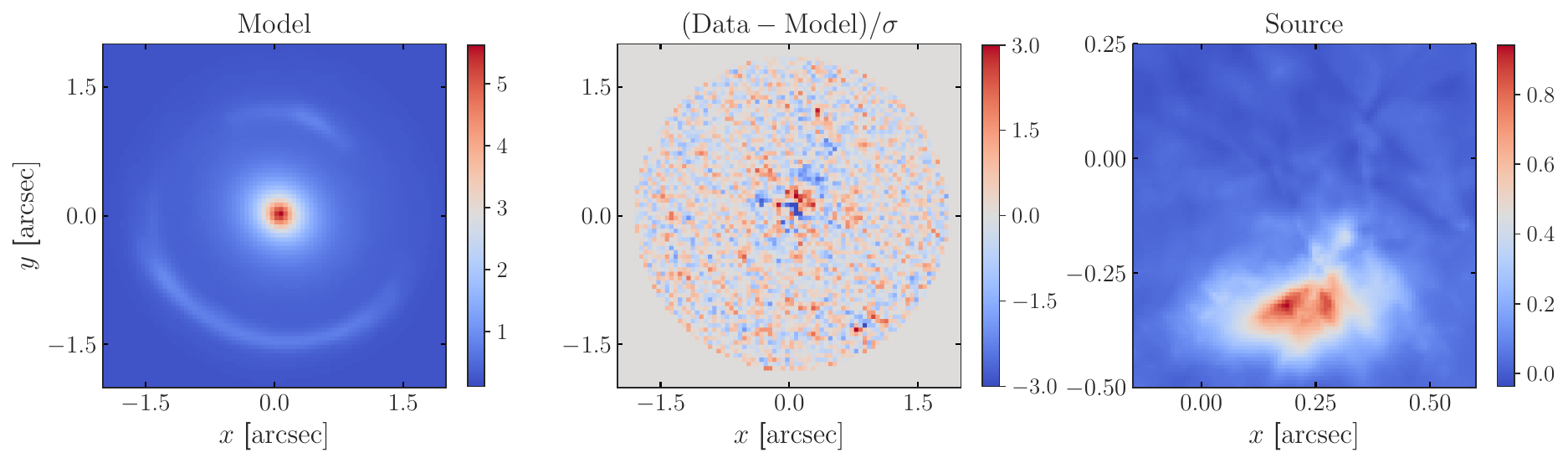}
    \caption{Same as \ref{fig:DR_RES}, but with supersampling performed on the image pixels.}
    \label{fig:SUPER_DR_NFW}
\end{figure*}

\begin{figure*}
\centering
    \includegraphics[width=\textwidth]{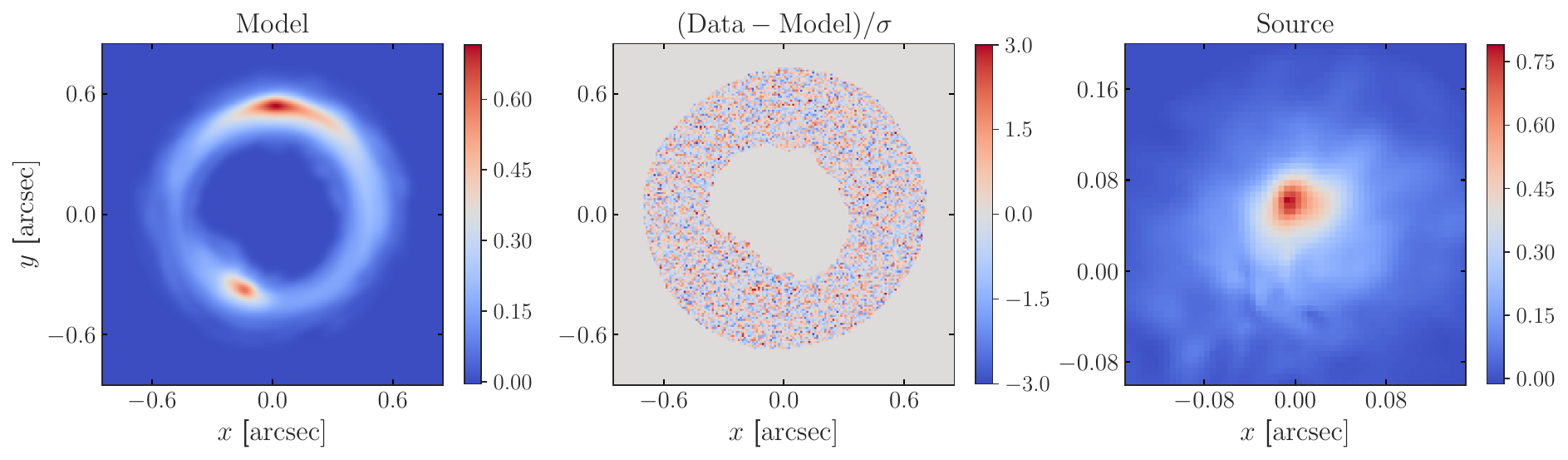}
    \caption{From left to right: the model, normalised image residuals and reconstructed source with the curvature regularisation, for the MAP parameters of the
best-fitting model that includes the PL field halo and multipoles for the gravitational lens system B1938+666.}
    \label{fig:B1938_RES}
\end{figure*}

\begin{figure*}
\centering
    \includegraphics[width=\textwidth]{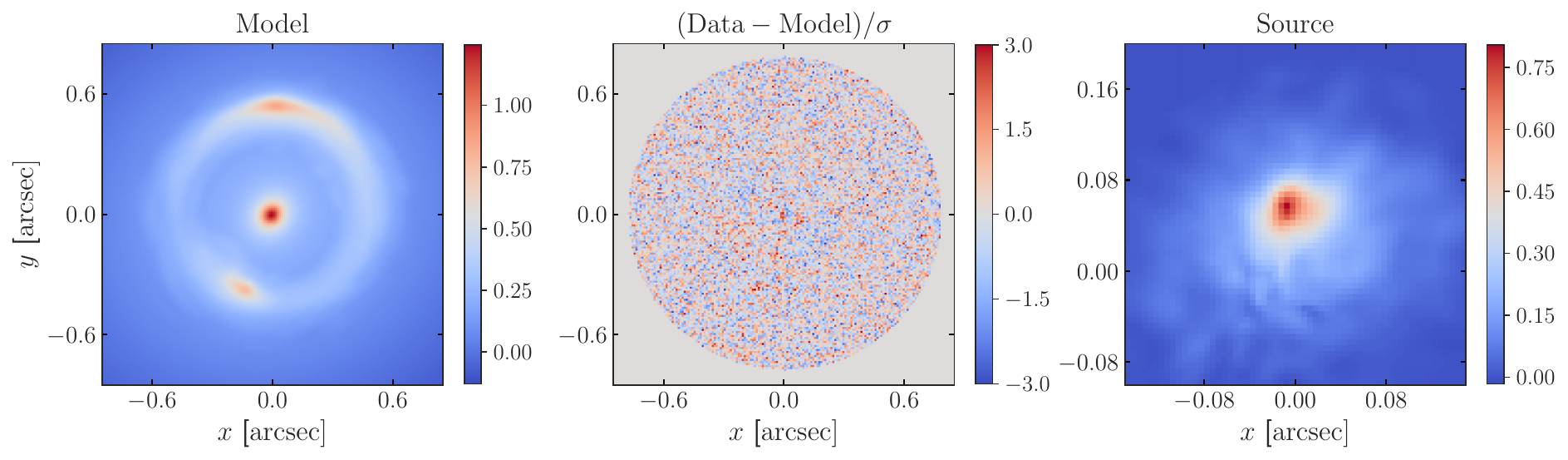}
    \caption{Same as \ref{fig:B1938_RES}, but for the case where the lens light and mass distributions are fitted simultaneously.}
    \label{fig:B1938_RES_LIGHT}
\end{figure*}

\begin{figure*}
\centering
    \includegraphics[width=\textwidth]{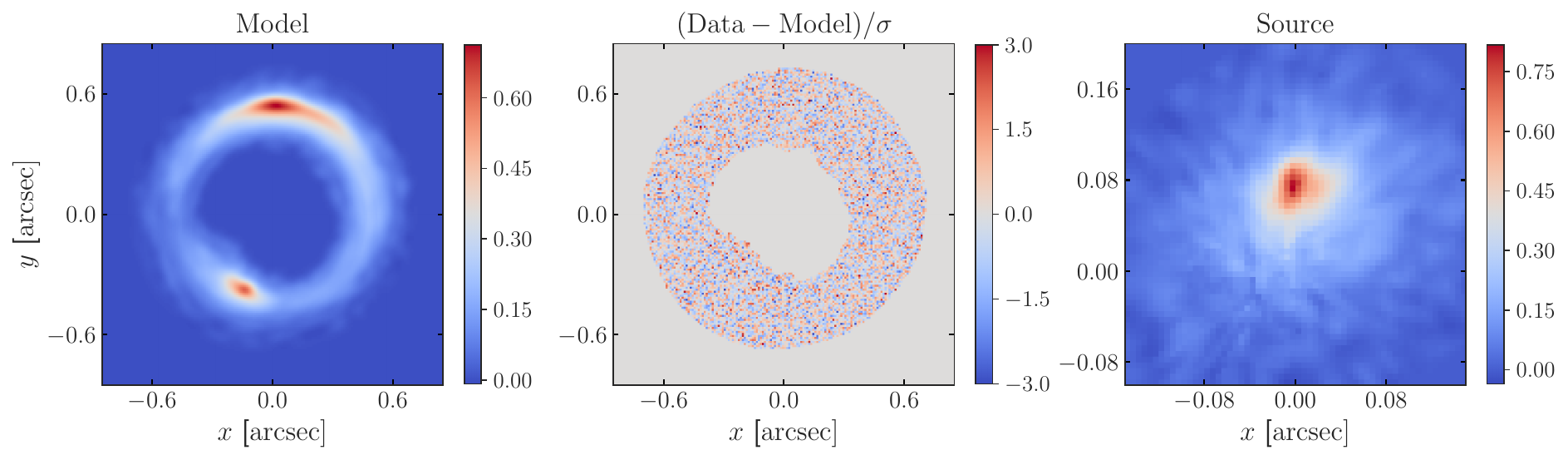}
    \caption{Same as \ref{fig:B1938_RES}, but with the gradient regularisation.}
    \label{fig:B1938_RES_GRAD}
\end{figure*}

Figures \ref{fig:macro_J0946} and \ref{fig:macro_b1938} show the posterior distributions of the main lens mass model parameters (power-law plus external shear and multipoles) for the best-fitting models of the gravitational lens systems J0946+1006 and B1938+666, respectively. The corresponding mass parameters and their uncertainties are listed in Table \ref{tab:mass_models}, and the foreground light parameters, modelled as a combination of Sérsic profiles, are provided in Table \ref{tab:light_models}. Figures \ref{fig:DR_mn_sub_vmax} to \ref{fig:DR_mn_pl} and \ref{fig:b1938_mn_pl} to \ref{fig:b1938_mn_duffy} show the posterior distributions of the parameters for the different mass models considered for detections $\cal{H}$ and $\cal{A}$, respectively. The reconstructed source and lensed emission, corresponding to the maximum-a-posteriori (MAP) parameters of the best-fitting models for each lens system, are presented in Figures \ref{fig:DR_RES} to \ref{fig:B1938_RES_GRAD}.

\begin{table*}
\caption{Mean inferred parameters of the mass density distribution of the main lens for both gravitational lens systems, based on the best-fitting models. Uncertainties correspond to the 68 per cent credible intervals.}
\label{tab:mass_models}
\centering
\begin{tabular}{lcc}
\toprule
 & J0946+1006 & B1938+666 \\
\midrule
$\kappa_0$ & $1.21 \pm 0.04$ & $0.54\pm0.02$ \\
$\theta$ [deg] & $-63 \pm 5$ &$-23\pm1$ \\
$q$ & $0.93 \pm 0.02$ & $0.82\pm0.01$ \\
$x$ [arcsec] & $0.03 \pm 0.01$ & $-0.027\pm0.001$ \\
$y$ [arcsec] & $0.05 \pm 0.01$ & $-0.024\pm0.005$ \\
$\alpha$ & $2.36 \pm 0.05$ & $1.86\pm0.03$ \\
$\Gamma$ & $0.098 \pm 0.004$ & $-0.033\pm0.004$ \\
$\theta_\Gamma$ [deg] & $-18 \pm 1$ & $-12\pm3$ \\
$\eta_1$ & $0.092 \pm 0.010$ & - \\
$\eta_3$ & $0.010 \pm 0.002$ & $0.002\pm0.001$ \\
$\eta_4$ & $0.011 \pm 0.002$ & $0.004\pm0.001$ \\
\bottomrule
\end{tabular}
\end{table*}

\begin{table*}
\caption{Mean Sérsic parameters for the surface brightness distribution of the main lens for both gravitational lens systems based on the best-fitting models. Uncertainties correspond to the 68 per cent credible intervals.}
\label{tab:light_models}
\small
\centering
\begin{tabular}{lccc|cc}
\toprule

& \multicolumn{3}{c|}{J0946+1006} 
& \multicolumn{2}{c}{B1938+666} \\
& 1 & 2 & 3 & 1 & 2 \\
\midrule
$I_{\rm e}$ [electrons s$^{-1}$ arcsec$^{-2}$] 
& $1.35\pm0.02$ & $0.90\pm0.04$ & $0.54\pm0.08$ 
& $0.64\pm0.09$ & $0.75\pm0.02$ \\
$R_{\rm e}$ [arcsec]         
& $0.66\pm0.02$ & $0.11\pm0.01$ & $0.15\pm0.02$ 
& $3.74\pm0.44$ & $0.081\pm0.002$ \\
$n_{\rm e}$                  
& $1.65\pm0.03$ & $0.98\pm0.05$ & $1.25\pm0.14$ 
& $0.71\pm0.03$ & $1.25\pm0.03$ \\
$x$ [arcsec]                 
& $0.041\pm0.001$ & $0.007\pm0.003$ & $0.116\pm0.004$ 
& $-0.017\pm0.002$ & $-0.0148\pm0.0003$ \\
$y$ [arcsec]                 
& $0.003\pm0.001$ & $-0.040\pm0.003$ & $0.127\pm0.004$ 
& $0.013\pm0.002$ & $-0.0066\pm0.0003$ \\
$\theta$ [deg]              
& $-124\pm1$ & $-129\pm1$ & $-127\pm2$ 
& $116\pm2$ & $143\pm2$ \\
$q$                          
& $0.868\pm0.004$ & $0.64\pm0.02$ & $0.54\pm0.02$ 
& $0.91\pm0.01$ & $0.86\pm0.01$ \\
\bottomrule
\end{tabular}
\end{table*}

\section{Fitting density profiles to simulated haloes}
\label{sec:profile_fitting}

\begin{figure*}
    \centering
	\includegraphics[width=1.0\textwidth]{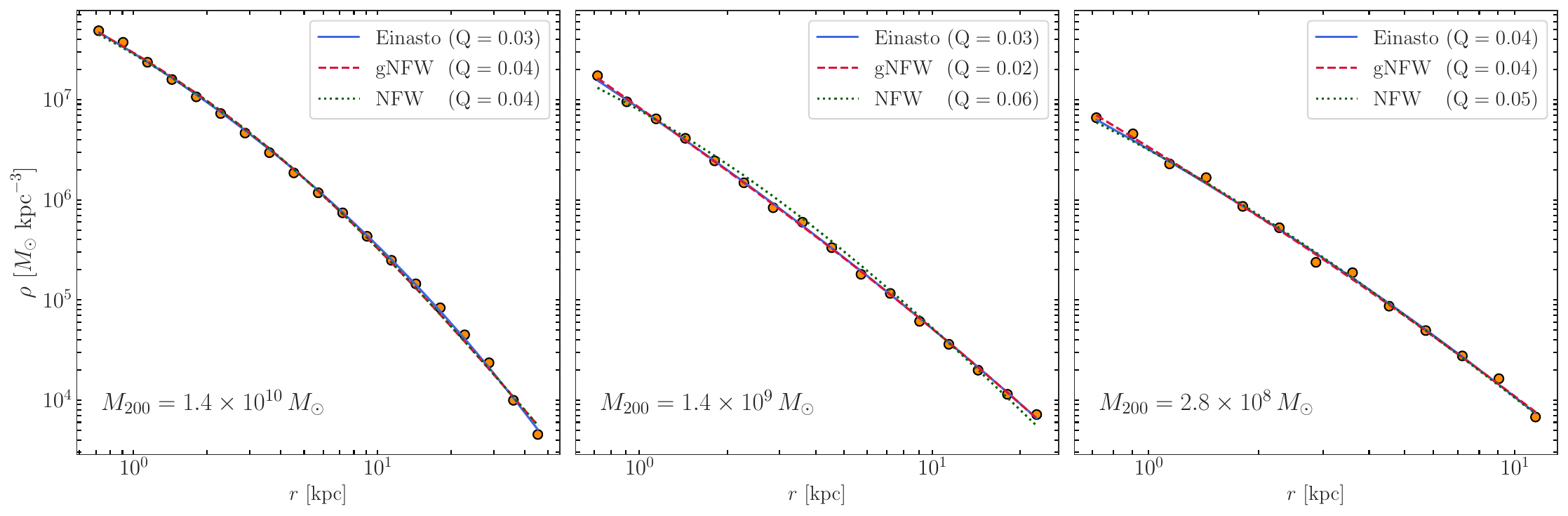}
    \caption{Mass density profile for three central haloes in the TNG50 simulation (orange circles) with redshift and total mass consistent with detection ${\cal A}$, and the corresponding best fits for an Einasto (blue lines), a generalised NFW (dashed red lines) and an NFW (dotted green lines) profile.}
    \label{fig:fit_haloes}
\end{figure*}

We fit the mass density distribution of the simulated haloes with three different models, an NFW (see Eq. \ref{eq:nfw}), an Einasto \citep{Einasto_1965} and a generalised NFW profile \citep{Zhao_1996}, with the latter two defined, respectively, as follows:
\begin{equation}
\label{einasto}
\rho(r) = \rho_s \exp \left(-\frac{2}{\alpha_{\rm e}} \left[\left(\frac{r}{r_s}\right)^{\alpha_{\rm e}} - 1\right] \right)\,,
\end{equation}
and
\begin{equation}
\label{gnfw}
\rho(r) = \frac{\rho_s}{(r/r_s)^\beta \left( 1 + r/r_s \right)^{3-\beta}}\,.
\end{equation}
To calculate the spherically averaged density profile, we define the halo centre as the position of the most bound particle (i.e. the one with the minimum gravitational potential) and compute the density in logarithmically spaced shells with $\Delta r = 0.1$ comoving kpc within the range $r \in [3\epsilon, R_{200, h}]$, where $R_{200, h}$ is the radius within which the mean density of the halo is 200 times the critical density of the Universe and varies for each halo. We note that fitting beyond $3\epsilon$ excludes the artificial core that forms at the centre due to limited resolution. Previous studies \citep[see, e.g.,][]{Springel_2008} have performed convergence tests on the density profiles of both haloes and subhaloes, demonstrating that, above $3\epsilon$, the profile slope is reliably recovered—even at radii smaller than the nominal resolution limit. All particle types (i.e. gas, stars, dark matter and black holes) are included in the profile calculation, regardless of whether they are identified as bound to the halo, to avoid biases introduced by specific subhalo finder algorithms. The best-fitting model parameters are obtained by minimising the following expression using the Levenberg-Marquardt method:
\begin{equation}
 Q^2 = \frac{1}{N_{\text{bin}}} \sum_{i=1}^{N_{\text{bin}}} \left( \log_{10} \rho_i - \log_{10} \rho_{i, \text{model}}\right)^2\,.  
\end{equation}
Here, $\rho_i$ is the measured density in radial bin $i$, and $\rho_{i,\text{model}}$ is the corresponding value predicted by the analytical model. This metric quantifies the goodness of fit, with lower $Q^2$ values indicating smaller deviations between the simulated density profile and the model. 
Fig. \ref{fig:fit_haloes} shows the results of this analysis for three simulated haloes of different masses, as an example.

\bsp	
\label{lastpage}
\end{document}